\documentclass[twocolumn]{aastex631}

\newcommand{\sm}{\,M$_{\odot}$}
\shorttitle{SN\,2020acct: A double-peaked PPISN candidate}
\shortauthors{Angus et al.}
\graphicspath{{./}{figures/}}

\begin{document}

\title{Double ``acct'': a distinct double-peaked supernova matching pulsational pair-instability models}

\author[0000-0002-4269-7999]{C.~R.~Angus}
\affiliation{Astrophysics Research Centre, School of Mathematics and Physics, Queen’s University Belfast, Belfast BT7 1NN, UK}
\affiliation{DARK, Niels Bohr Institute, University of Copenhagen, Jagtvej 128, DK-2200 Copenhagen {\O} Denmark}

\author[0000-0002-3352-7437]{S.~ E.~Woosley}
\affiliation{Department of Astronomy and Astrophysics, University of California, Santa Cruz, CA 95064, USA}
\author[0000-0002-2445-5275]{R.~J.~Foley}
\affiliation{Department of Astronomy and Astrophysics, University of California, Santa Cruz, CA 95064, USA}
\author[0000-0002-2555-3192]{M.~Nicholl}
\affiliation{Astrophysics Research Centre, School of Mathematics and Physics, Queen’s University Belfast, Belfast BT7 1NN, UK}
\author[0000-0002-5723-8023]{V.~A.~Villar}
\affiliation{Center for Astrophysics \textbar{} Harvard \& Smithsonian, Cambridge, MA 02138, USA}
\affiliation{The NSF AI Institute for Artificial Intelligence and Fundamental Interactions}
\author[0000-0002-5748-4558]{K.~Taggart}
\affiliation{Department of Astronomy and Astrophysics, University of California, Santa Cruz, CA 95064, USA}
\author[0000-0003-4663-4300]{M.~Pursiainen}
\affiliation{Department of Physics, University of Warwick, Gibbet Hill Road, Coventry, CV4 7AL, UK}
\author[0009-0009-2627-2884]{P.~Ramsden}
\affiliation{School of Physics and Astronomy, University of Birmingham, Birmingham B15 2TT, UK}
\affiliation{Astrophysics Research Centre, School of Mathematics and Physics, Queen’s University Belfast, Belfast BT7 1NN, UK}
\author[0000-0003-4524-6883]{S. Srivastav}
\affiliation{Astrophysics, Department of Physics, University of Oxford, Keble Road, Oxford, OX1 3RH, UK}
\author[0000-0002-0504-4323]{H.~F.~Stevance}
\affiliation{Astrophysics, Department of Physics, University of Oxford, Keble Road, Oxford, OX1 3RH, UK}
\affiliation{Astrophysics Research Centre, School of Mathematics and Physics, Queen’s University Belfast, Belfast BT7 1NN, UK}
\author[0000-0002-6842-3021]{T.~Moore}
\affiliation{Astrophysics Research Centre, School of Mathematics and Physics, Queen’s University Belfast, Belfast BT7 1NN, UK}
\affiliation{European Southern Observatory, Alonso de C\'{o}rdova 3107, Casilla 19, Santiago, Chile}
\author[0000-0002-4449-9152]{K.~Auchettl}
\affiliation{School of Physics, The University of Melbourne, VIC 3010, Australia}
\affiliation{Department of Astronomy and Astrophysics, University of California, Santa Cruz, CA 95064, USA}
\author[0000-0003-3953-9532]{W.~B.~Hoogendam}
\affiliation{Institute for Astronomy, University of Hawaii, 2680 Woodlawn Drive, Honolulu, HI 96822, USA}
\author[0000-0003-2720-8904]{N.~Khetan}
\affiliation{School of Mathematics and Physics, The University of Queensland, QLD 4072, Australia}
\author[0000-0002-0840-6940]{S.~K.~Yadavalli}
\affiliation{Center for Astrophysics \textbar{} Harvard \& Smithsonian, Cambridge, MA 02138, USA}
\author[0000-0001-9494-179X]{G.~Dimitriadis}
\affiliation{Department of Physics, Lancaster University, Lancs LA1 4YB, UK}
\author[0000-0003-4906-8447]{A.~Gagliano}
\affiliation{The NSF AI Institute for Artificial Intelligence and Fundamental Interactions}
\affiliation{Center for Astrophysics \textbar{} Harvard \& Smithsonian, Cambridge, MA 02138, USA}
\affiliation{Department of Physics, Massachusetts Institute of Technology, Cambridge, MA 02139, USA}
\author[0000-0003-2445-3891]{M.~R.~Siebert}
\affiliation{Space Telescope Science Institute, Baltimore, MD 21218, USA}
\author[0000-0002-9085-8187]{A.~Aamer}
\affiliation{Astrophysics Research Centre, School of Mathematics and Physics, Queen’s University Belfast, Belfast BT7 1NN, UK}
\author[0000-0001-5486-2747]{T.~de~Boer}
\affiliation{Institute for Astronomy, University of Hawaii, 2680 Woodlawn Drive, Honolulu, HI 96822, USA}
\author[0000-0001-6965-7789]{K.~C.~Chambers}
\affiliation{Institute for Astronomy, University of Hawaii, 2680 Woodlawn Drive, Honolulu, HI 96822, USA}
\author[0000-0003-3068-4258]{A.~ Clocchiatti}
\affiliation{Millennium Institute of Astrophysics (MAS), Nuncio Monse\~{n}or S\'{o}tero Sanz 100, Of. 104, Providencia, Santiago, Chile}
\affiliation{Instituto de Astrof\'{i}sica, Facultad de F\'{i}sica, Pontificia Universidad Cat\'{o}lica de Chile, Vicu\~{n}a Mackenna 4860, Macul, Santiago, Chile}
\author[0000-0003-4263-2228]{D.~A.~Coulter}
\affiliation{Space Telescope Science Institute, Baltimore, MD 21218, USA}
\author[0000-0001-7081-0082]{M.~R.~Drout}
\affiliation{David A. Dunlap Department of Astronomy and Astrophysics, University of Toronto, 50 St. George Street, Toronto, Ontario, M5S 3H4, Canada}
\author[0000-0002-0786-7307]{D.~Farias}
\affiliation{DARK, Niels Bohr Institute, University of Copenhagen, Jagtvej 128, DK-2200 Copenhagen {\O} Denmark}
\author[0000-0003-1916-0664]{M.~D.~Fulton}
\affiliation{Astrophysics Research Centre, School of Mathematics and Physics, Queen’s University Belfast, Belfast BT7 1NN, UK}
\author[0000-0002-8526-3963]{C.~Gall}
\affiliation{DARK, Niels Bohr Institute, University of Copenhagen, Jagtvej 128, DK-2200 Copenhagen {\O} Denmark}
\author[0000-0003-1015-5367]{H.~Gao}
\affiliation{Institute for Astronomy, University of Hawaii, 2680 Woodlawn Drive, Honolulu, HI 96822, USA}
\author[0000-0001-9695-8472]{L.~Izzo}
\affiliation{INAF-Osservatorio Astronomico di Capodimonte, Salita Moiariello 16, 80131 Naples, Italy}
\affiliation{DARK, Niels Bohr Institute, University of Copenhagen, Jagtvej 128, DK-2200 Copenhagen {\O} Denmark}
\author[0000-0002-6230-0151]{D.~O.~Jones}
\affiliation{Institute for Astronomy, University of Hawaii, 640 N A`ohoku Pl, Hilo, Hawai`i, USA}
\author[0000-0002-7272-5129]{C.-C.~Lin}
\affiliation{Institute for Astronomy, University of Hawaii, 2680 Woodlawn Drive, Honolulu, HI 96822, USA}
\author[0000-0002-7965-2815]{E.~A.~Magnier}
\affiliation{Institute for Astronomy, University of Hawaii, 2680 Woodlawn Drive, Honolulu, HI 96822, USA}
\author[0000-0001-6022-0484]{G.~Narayan}
\affiliation{Department of Astronomy, University of Illinois at Urbana-Champaign, 1002 W. Green St., IL 61801, USA}
\affiliation{Center for Astrophysical Surveys, National Center for Supercomputing Applications, Urbana, IL, 61801, USA}
\author[0000-0003-2558-3102]{E.~Ramirez-Ruiz}
\affiliation{Department of Astronomy and Astrophysics, University of California, Santa Cruz, CA 95064, USA}
\author[0000-0003-4175-4960]{C.~L.~Ransome}
\affiliation{Center for Astrophysics \textbar{} Harvard \& Smithsonian, Cambridge, MA 02138, USA}
\author[0000-0002-4410-5387]{A.~Rest}
\affiliation{Department of Physics and Astronomy, The Johns Hopkins University, Baltimore, MD 21218, USA}
\affiliation{Space Telescope Science Institute, Baltimore, MD 21218, USA}
\author[0000-0002-8229-1731]{S.~J.~Smartt}
\affiliation{Astrophysics, Department of Physics, University of Oxford, Keble Road, Oxford, OX1 3RH, UK}
\affiliation{Astrophysics Research Centre, School of Mathematics and Physics, Queen’s University Belfast, Belfast BT7 1NN, UK}
\author[0000-0001-9535-3199]{K.~W.~ Smith}
\affiliation{Astrophysics Research Centre, School of Mathematics and Physics, Queen’s University Belfast, Belfast BT7 1NN, UK}




\begin{abstract}
We present multi-wavelength data of SN\,2020acct, a double-peaked stripped-envelope supernova (SN) in NGC\,2981 at $\sim$150~Mpc. The two peaks are temporally distinct, with maxima separated by 58 rest-frame days, and a factor of 20 reduction in flux between. The first is luminous ($M_{r}=-18.00\pm0.02$~mag), blue ($g-r=0.27\pm0.03$~mag), and displays spectroscopic signatures of interaction with hydrogen-free circumstellar material. The second peak is fainter ($M_{r}=-17.29\pm0.03$~mag), and spectroscopically similar to an evolved stripped-envelope SNe, with strong blended forbidden [\ion{Ca}{2}] and [\ion{O}{2}] features. No other known double-peak SN exhibits a light curve similar to that of SN\,2020acct. We find the likelihood of two individual SNe occurring in the same star-forming region within that time to be highly improbable, while an implausibly fine-tuned configuration would be required to produce two SNe from a single binary system. We find that the peculiar properties of SN\,2020acct match models of pulsational pair instability (PPI), in which the initial peak is produced by collisions of shells of ejected material, shortly followed by a terminal explosion. Pulsations from a star with a 72\,M$_{\odot}$ helium core provide an excellent match to the double-peaked light curve. The local galactic environment has a metallicity of $0.4Z_{\odot}$, a level where massive single stars are not expected retain enough mass to encounter the PPI. However, late binary mergers or a low-metallicity pocket may allow the required core mass. We measure the rate of SN\,2020acct-like events to be $<${}$3.3\times10^{-8}$~Mpc$^{-3}$\,yr$^{-1}$ at $z=0.07$, or $<$0.1\% of the total core-collapse SN rate.
\end{abstract}



\section{Introduction} \label{sec:intro}

The observed diversity of the supernova (SN) population has expanded in recent years as a result of the increased depth, wavelength coverage, and observing baselines afforded by modern surveys. This is most apparent within core-collapse SNe (CCSNe), the final explosions of stars $\gtrsim8$\sm. Though CCSNe are still spectroscopically categorized as hydrogen-rich (type II), hydrogen-deficient (type IIb), and hydrogen-poor (types Ib/Ic), their increasingly heterogeneous light curve morphologies raise questions about the similarity of their progenitor systems, explosion mechanisms, and our understanding of massive stellar evolution. 

In recent years, a growing number of ``double-peaked" SNe have been identified, whose light curves exhibit a secondary maximum either preceding or trailing the main SN. Double peaks have been observed within both hydrogen-rich and hydrogen-poor SN classes, and the properties of the second peaks are themselves diverse. Strong rebrightenings post-SN have been attributed to a variety of physical mechanisms, including internal engines in the form of accretion onto a newly formed compact object \citep{Moore2023,Chen2024,Kangas2024}, spin down of a newly formed magnetar \citep{Maeda2007,Gutierrez2021}, the interaction of the SN ejecta with material flung far from the progenitor \citep{Sollerman2020,Sharma2024}, or even double-peaked $^{56}$Ni distributions in the ejecta \citep{Taddia2018,Tominaga2005}. This is in contrast to peaks which occur {\textit{before}} the SN, which are usually attributed to the circumstellar interaction (CSI) of material lost from the progenitor star in the months-years leading up to its explosion \citep[although see][for alternative models]{Kasen2016}.

More common early emission signatures range from the bright and rapidly declining shock-cooling peaks in some SNe~IIb, \citep[e.g.][]{Waxman&Katz2017}, and the faint, brief early bumps observed in some hydrogen-poor superluminous SNe \citep[e.g.][]{Leloudas2012,Nicholl2016a,Angus2019,Chen2023}. These ``pre-peaks" occur immediately prior to the rise of the main SN, and are thought to arise from shock-cooling of circumstellar material (CSM) \citep{Nicholl2015,Piro2015,Smith2016} close to the progenitor star. However, other early peaks show a greater separation from the main SN, occurring in the weeks-months prior to the main explosion. Such ``precursor" events are thought to be the result of late mass ejection from the progenitor star. While smaller precursor eruptions are believed to be common in red supergiants \citep{Jacobson-Galan2022}, there is growing evidence that massive stars undergo large-scale eruptions, shedding up to $\sim$few$\,M_{\odot}$ of material in a single episode \citep{Smith2003}. These eruptions can result in luminous transient events, capable of achieving SN-like luminosities \citep{Mauerhan2013,Fraser2013,Margutti2014,Graham2014}. Pre-explosion outbursts have been witnessed in a growing sample of SN progenitors in the weeks to years prior to their final collapse \citep[e.g][]{Fraser2013,Ofek2013,Pastorello2015,Brennan2024}, and a wide variety of scenarios have been invoked to explain them; from binary interaction \citep{Yoon2010}, to Luminous Blue Variable outbursts \citep{HumphreysDavidson1994}.

Pulsational Pair Instability (PPI) lies at the extreme end of eruptive mass-loss episodes. In stars with Zero Age Main Sequence (ZAMS) masses 70$\,\lesssim$\,M\,$\lesssim$\,140\,M$_{\odot}$, their massive ($\gtrsim35\,M_{\odot}$) carbon and oxygen-rich cores can reach temperatures of $\gtrsim5\times10^{8}$\,K, leading to the production of electron-positron pairs,  removing pressure support from the core \citep{Rakavy1967}. This leads to a dynamically unstable contraction of the O-rich core, which ignites explosive O-burning. The energy released is insufficient to disrupt the whole star, but instead generates strong pulsations, driving mass ejections from the surface \citep{Rakavy1967,Barkat1967}. Such pulses are capable of expelling $\sim10$\,M$_{\odot}$ worth of material from the outer envelope of the progenitor \citep{Woosley2017,Renzo2020}, and a star may undergo multiple PPI ejections throughout its lifetime, ceasing once the entropy loss from pulsations alleviates the instability. Collisions between successively ejected shells can produce a bright, SN-like transient in the months-years prior to the final explosion \citep{Woosley2007,Woosley2017}. 

PPI events almost certainly exist in the Universe, as even at moderate ($\sim$solar) metallicity, there will be a mass range of helium and heavy element cores that experience PPI, produce violent mass ejections, and eventually collapse to form heavy stellar mass black holes \citep[$40-50\,M_{\odot}$][]{Barkat1967,Woosley2017}. The detection of massive black hole binaries by the LIGO-Virgo collaboration \citep{Abbott2023,LigoVirgo2023} is indicative that such events have occurred. Due to the structured CSM with multiple shells that PPI is predicted to produce, PPI models can physically describe SNe whose light curves show peculiar morphologies (e.g. significant pre-explosion excesses, large modulations or bumps in the light curve, or long-lasting blue and luminous events), or those whose spectra exhibit signs of strong interaction. Observable transients from PPI (``PPISNe'') may result from collisions between shells of material ejected close in time, or from the collision of the ejecta from a terminal SN with previously expelled shells, allowing them to span a wide range of luminosities and durations. 

However, observational confirmation of PPI events is difficult, as other mass-loss mechanisms can also produce strongly-interacting SNe. The unique identifier of a PPI event is the presence of an interaction-powered transient with SN-like energies prior to the actual explosion of the star. Several notable PPISN candidates have been identified within recent years, including: SN\,2010mb \citep{BenAmi2014}, iPTF14hls \citep{Arcavi2017}, SN\,2017egm \citep{Lin2023,Zhu2023}, iPTF16eh \citep{Lunnan2018}, SN\,2016aps \citep{Nicholl2020}, SN\,2016eit \citep{Gomez2019} and SN\,2019szu \citep{Aamer2023}. Though these events show evidence of a CSM shell that is detached or massive (iPTF16eh, SN\,2016aps), and perhaps complex in structure (SN\,2010mb, iPTF14hls, SN\,2017egm), none are unambiguous PPISN events, either because there are no detections of the pre-explosion shell collisions that uniquely identify the mass-loss mechanism, or because any early peaks lack the spectroscopic or photometric coverage to confidently say that the first peak is both interaction-driven (SN\,2019szu) and occurs before final core-collapse (iPTF16eit). Moreover, none of these events have been matched to a specific PPISN simulation that can reproduce the observables.

In this paper, we present SN\,2020acct, a double-peaked SN whose peaks are photometrically and spectroscopically distinct and whose properties point toward a PPI origin. Throughout this work we assume a standard $\Lambda$CDM cosmology of H$_{0}$=70\,km\,s$^{-1}$\,Mpc$^{-1}$, $\Omega_{M}$=0.3 and $\Omega_{\Lambda}$=0.7. 

\section{Data} \label{sec:data}
\subsection{Photometry} \label{sec:photdata}
\begin{figure*}
\includegraphics[width=\textwidth]{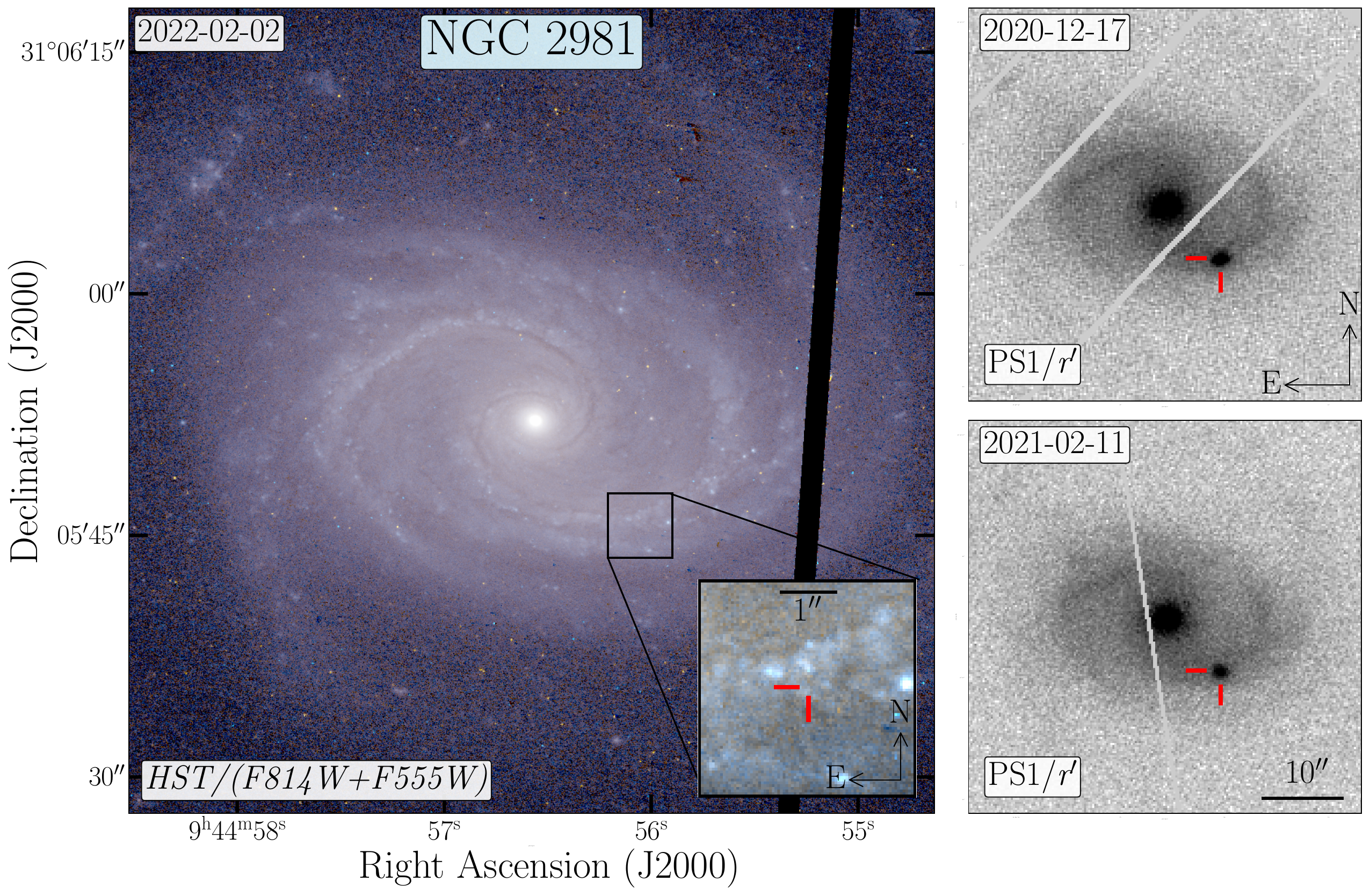}
\caption{Images of NGC\,2981, the host galaxy of SN\,2020acct. {\textit{Left}}: Stacked {\it HST} images (F275W, F555W and F814W bands). The 3\arcsec$\times$3\arcsec\ inset image zoomsin on the local  environment with red crosshairs marking the location of SN\,2020acct. {\textit{Right}}: PanSTARRS-1 $r$-band images of SN\,2020acct during the first (\textit{top}) and second (\textit{bottom}) peaks used for astrometric measurements. }
\label{fig:host}
\end{figure*}

SN\,2020acct (ZTF20acwobku) was discovered by the Zwicky Transient Facility \citep[ZTF;][]{Bellm2019} in the $g$-band on 2020 December 10 16.75 UT, or MJD 59193.923, at RA=09:44:56.060, $\delta$=+31:05:45.77. Located in NGC\,2981 at $z=0.034650$ \citep{SDSSDR13}, SN\,2020acct is offset 6.24''S, 6.27''W from the core of the galaxy (a physical separation of 6.1 kpc). The location of SN\,2020acct within NGC\,2981 is shown in Figure \ref{fig:host}.  The field of SN\,2020acct was also observed as part of routine operations by the Young Supernova Experiment \citep[YSE;][]{YSEsurvey}, the Pan-STARRS survey for Transients \citep[PSST; ][]{PanSTARRSsurvey} and the Asteroid Terrestrial-impact Last Alert System \citep[ATLAS;][]{ATLASsurvey}. The YSE-PZ web broker was used for coordinating followup observations and visualisation of the YSE, ATLAS and public ZTF data \citep{Coulter2022,Coulter2023}.

$g$,$r$,$i$,$z$-band imaging was obtained through YSE with the Pan-STARRS telescope \citep[PS;][]{PSTelescopes} between 2020 December 09 and 2021 February 22. All images were reduced with the YSE photometric pipeline, which is based on {\texttt{photpipe}} \citep{Rest2005}, with template images taken from stacked Pan-STARRS1 $3\pi$ sky survey data \citep[][Fulton et al. {\textit{in prep.}}]{PanSTARRSsurvey}. All images and templates are resampled and astrometrically aligned to match a skycell in the PS1 sky tessellation, and nightly image zero-points determined by comparing PSF photometry to PS1 stellar catalogues \citep{PanSTARRSsurvey}. PSF-matched templates are subtracted from the nightly images, and the flux-weighted centroid is determined at a position forced to match the SN position determined from the first images, from which we perform PSF photometry. Additional $i$,$z$,$y$-band Pan-STARRS imaging was obtained as part of PSST, alongside the composite $w$-band. PSST images were processed in real-time as described in \cite{Magnier2020b}. We perform forced photometry at the location of the first Pan-STARRS detection using the Pan-STARRS1 $3\pi$ sky survey data \citep{PanSTARRSsurvey} as our reference.

We use the ZTF forced photometry server \citep{Masci2023} to obtain public $g$- and $r$-band photometry of SN\,2020acct, with detections spanning 2020 December 10 to 2021 April 02 ($-$58 to +50\,d in the rest-frame). We use the ATLAS forced photometry server \citep{ATLASsurvey,Smith2020,Shingles2021} to recover the difference-image photometry in the $o$- and $c$-bands for SN\,2020acct, with detections covering the period 2020 December 10 to 2021 February 22. 

We obtained target-of-opportunity observations from the Neil Gehrels Swift Gamma-ray Burst Mission \citep[{\textit{Swift}};][]{Gehrels2004} UltraViolet and Optical Telescope \citep[UVOT;][]{Roming2005}. SN\,2020acct was monitored with {\textit{Swift}} from 2020 December 14 to 2020 December 23, and from 2021 February 12 to 2021 February 23. To estimate the host-galaxy level for subtraction, UVOT observations were also obtained after the SN had faded on 2021 October 11. We perform photometry using the HEAsoft package {\tt{UVOTSOURCE}} with a source aperture of 2.0'' to minimize contamination from surrounding regions of star formation and a source free background region with a radius of 20.0''.

Observations with the Hubble Space Telescope ({\it HST}) WFC3/UVIS of SN\,2020acct were obtained in the F275W band on 2021 November 29 (program GO--16657, PI: Fremling), and in the F555W and F814W bands on 2022 February 02 (program SNAP--16691 , PI: Foley). SN\,2020acct is undetected at both epochs, with 5$\sigma$-limits of $m_{\mathrm{275}}>28.3$, $m_{\mathrm{555}}>27.8$ and $m_{\mathrm{814}}>27.8$ within a 1'' aperture. 

\begin{figure*}
\includegraphics[width=\textwidth]{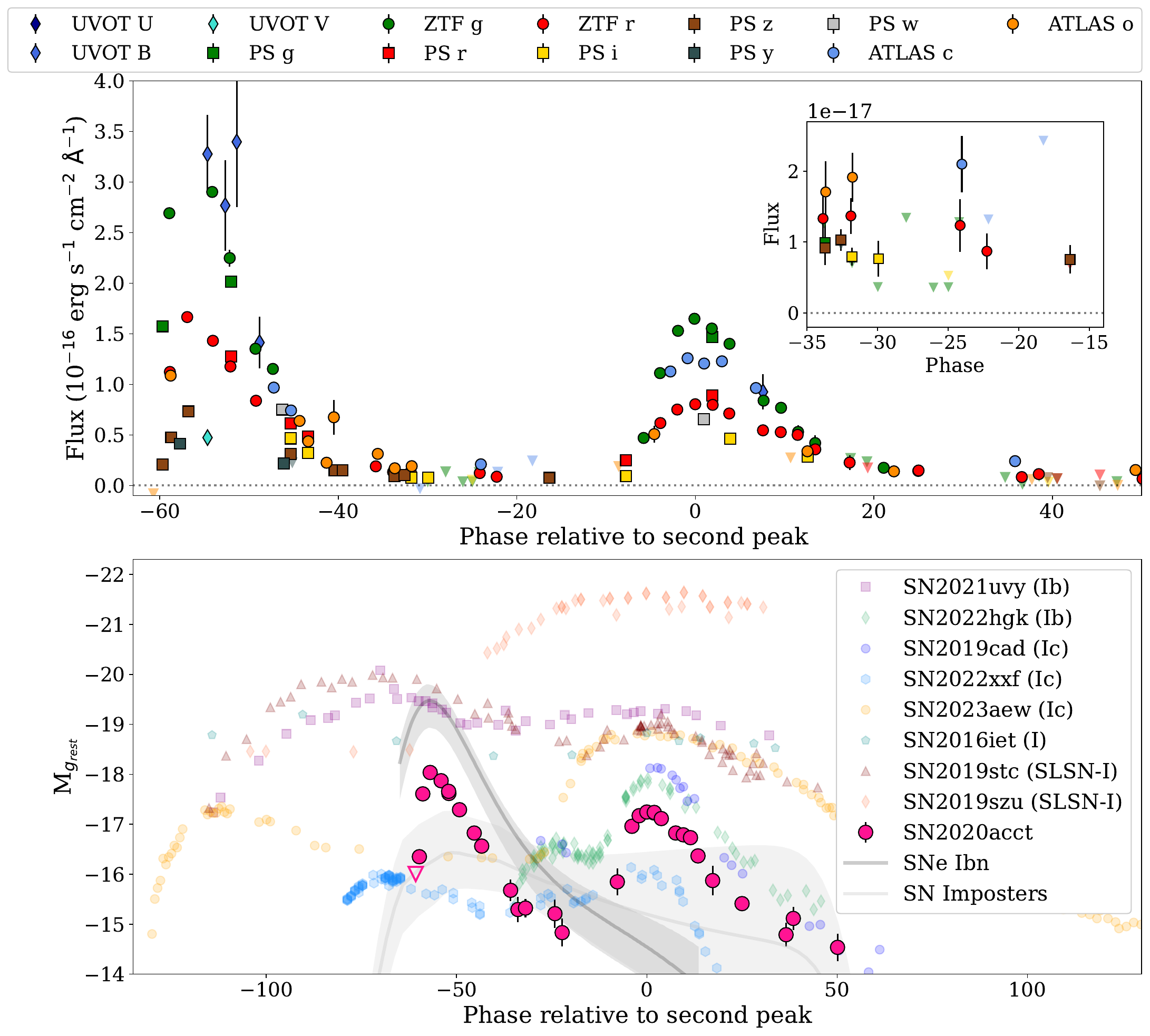}
\caption{{\textit{Top:}} The light curve of SN\,2020acct. The phase is given in the rest frame with respect to the maxima of the second peak (MJD 59254.4). The two peaks are distinct, separated by 58 days in the rest frame (maximum to maximum).  The inset shows that between the peaks the SN does not completely fade to zero flux. {\textit{Bottom:}} Photometric comparison of SN\,2020acct with other notable double peaked SESNe \citep{Gutierrez2021,Kuncarayakti2023,Das2023,Aamer2023,Kangas2024,Sharma2024}, including the PPISN candidates SN\,2016eit SN\,2019stc and SN\,2019szu \citep{Gomez2019,Gomez2021,Aamer2023}. We also show the \cite{Hosseinzadeh2017} SN~Ibn template light curve and a template light curve for SN impostors (derived using the light curves of \citealt{Fraser2015,Elias-Rosa2016,Tartaglia2016,Pastorello2018,Brennan2022}).  }
\label{fig:lc}
\end{figure*}

We estimate the time of explosion, $t_0=59192.1\pm0.5$, as the midpoint between the last pre-explosion non-detection from ATLAS on MJD 59191.58 and an early first detection from the YSE survey in the $g$- and $z$-bands on MJD 59192.65. We present the light curve of SN\,2020acct in the top panel of Figure \ref{fig:lc}, with detections spanning 106 rest-frame days from initial discovery. Photometric observations are provided within Appendix Table \ref{tab:phot}, where all rest-frame phases are reported with respect to the maximum of the second peak. All photometry is corrected for the Milky Way foreground extinction using the dust maps of \cite{SchlaflyFinkbeiner2011}. We do not correct for internal host galaxy extinction. 

\subsubsection{Astrometry} \label{sec:astrom}

We analyzed PS1 $r$-band images corresponding to the first and second peaks of SN\,2020acct, taken on 2020 December 17 and 2021 February 11, respectively, to determine the relative astrometric offset and confirm the coincidence of the two peaks. The zoomed-in images are displayed in the right panels of Figure~\ref{fig:host}.

Using {\tt Source Extractor} \citep{SourceExtractor}, we identified common stars in both images. We applied quality control cuts to exclude stars based on source ellipticity, the full width half maximum (FWHM), saturation, or proximity to the image edges. After removing these sources, the images had 30 stars in common with well-measured positions. 

To compute the geometric transformation between the two images, we performed iterative fitting on these stars using {\tt GEOMAP} \citep{IRAF}. Following the removal of 4 additional stars, we applied a second-order Legendre polynomial fit and used {\tt GEOTRAN} to apply this transformation. We performed additional systematic tests, including a jackknife test with the selected stars to determine the influence of individual stars in the fitting process and, by changing the order of the fit, finding an additional 0.07 and 0.07-pixel uncertainty in each direction, respectively.  

We find the location of the two peaks to be coincident, with a relative offset of (0.19 $\pm$ 1.71, 0.36 $\pm$ 1.38) pixels in the $x$ and $y$ dimensions, respectively, or (0.76 $\pm$ 6.84, 1.44 $\pm$ 5.52) arcsec,  indicating they are consistent to within 0.19-sigma.


\subsubsection{Long-term Light Curve}\label{sec:baseline}

We search for variability prior to the detection of the first peak of SN\,2020acct, performing forced photometry at its astrometric location. We use the ZTF \citep{Masci2019}, and ATLAS \citep{Smith2020} forced photometry servers to recover the difference-image photometry, and internally collected forced photometry for PSST, with observations dating from 01 Jan 2018 (ATLAS), 18 Nov 2013 (PSST) and 13 Nov 2019 (ZTF). We visually inspect the forced light curves and do not find significant pre-explosion activity within the PSST and ZTF data down to a luminosity of M$=-14.9$. For the shallower ATLAS data, we compare the forced light curve of SN\,2020acct to the forced photometry of eight control light curves located within a circular radius of 15' of the transient, using the weighted mean of the control light curves to identify any discrepant epochs in the target light curve \citep[see ][ for further details]{Rest2024}. We find no significant variability in the pre-explosion data out to MJD 57754 (01 Jan 2017), down to a luminosity of M$_{o}=-15.9$. 

\subsection{Spectroscopy}\label{sec:specdata}

We obtained 5 epochs of optical spectroscopic observations of SN\,2020acct spanning $-$58\,d to +56\,d (rest-frame). The optical spectra were obtained with the Kast dual-beam spectrograph \citep{Miller1993} on the Lick Shane telescope, the Alhambra Faint Object Spectrograph (ALFOSC) on the Nordic Optical Telescope (NOT), and the Low-Resolution Imaging Spectrograph \citep[LRIS;][]{Oke1995} on the Keck I telescope. A final spectrum was obtained using LRIS on 2021 May 12 (+89\,d), with the slit centred on the SN location after optical light from the second peak had faded. We use this spectrum for host galaxy corrections. A log of all spectroscopic observations and observing setups are presented in the Appendix.

\begin{figure*}
\includegraphics[width=\textwidth]{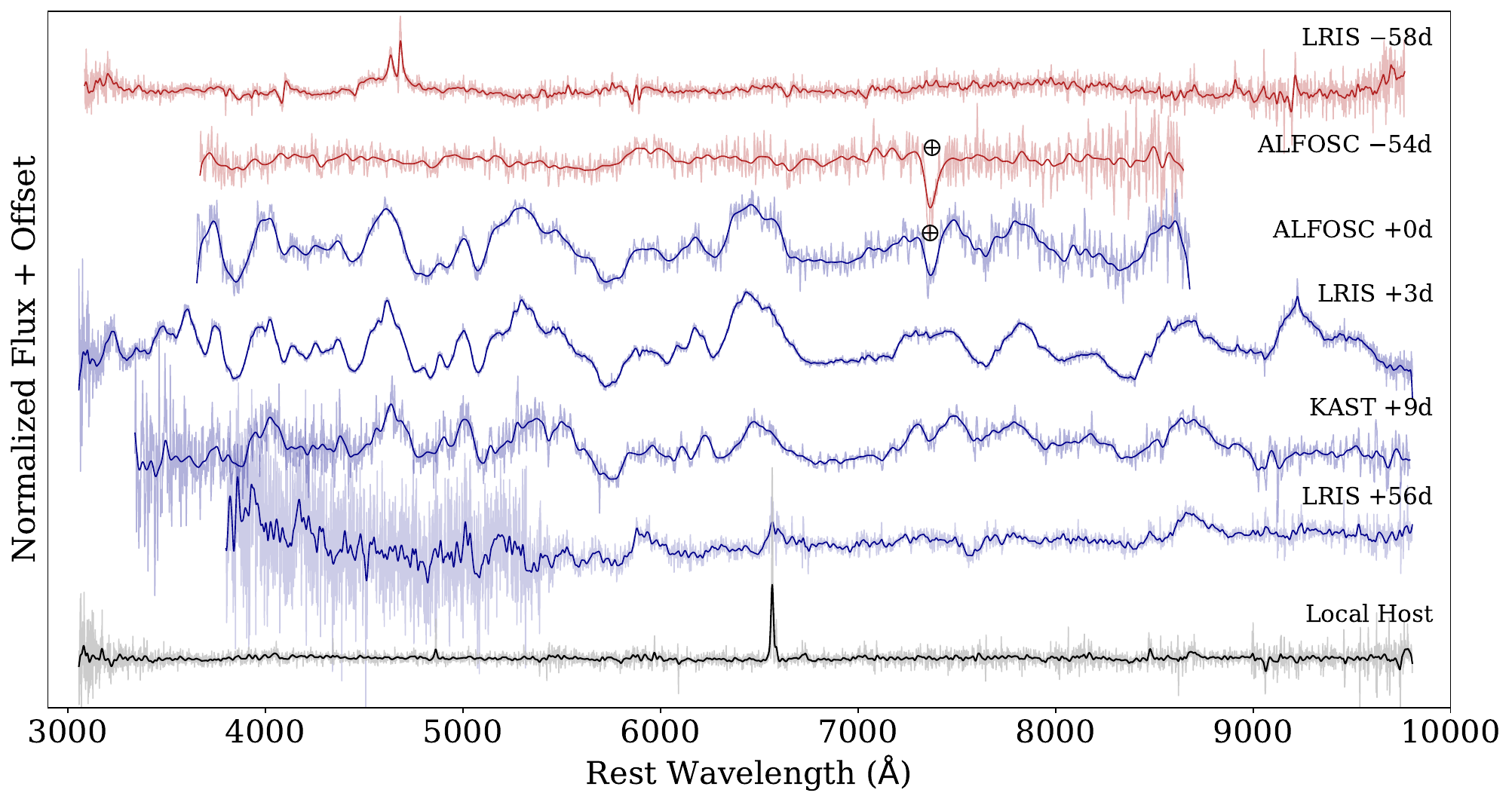}
\caption{Spectroscopic time series of SN\,2020acct. Spectra obtained during the first peak are shown in purple, while those obtained during the second peak are shown in turquoise. All phases are given with respect to the second peak. A host galaxy spectrum from the explosion site of SN\,2020acct is shown in black. The two peaks are spectroscopically distinct, with the first showing swiftly evolving narrow lines and the second showing absorption features consistent with a stripped-envelope SN. }
\label{fig:specev}
\end{figure*}

The LRIS and Kast spectra were reduced using the UCSC Spectral Pipeline\footnote{{\href{https://github.com/msiebert1/UCSC_spectral_pipeline}{https:\/\/github.com\/msiebert1\/UCSC\_spectral\_pipeline}}} and processed as described in \cite{Siebert2019}. The NOT-ALFOSC spectra were reduced using custom built {\tt{iraf}} pipelines. We color-correct all transient and host-galaxy spectra, mangling the spectra \citep{Hsiao2007} to the available photometry to correct for slit losses. We use the LRIS local host galaxy spectrum to perform our host-subtraction, convolving the host spectrum to lower resolution instruments where applicable before subtraction. We present the reduced spectra of SN\,2020acct in Figure \ref{fig:specev}. 

\section{Analysis} 

Below we examine the spectroscopic, photometric and host-galaxy properties of SN\,2020acct, outlining its physical characteristics in the context of other CCSNe.

\subsection{Host Galaxy Environment} \label{sec:host}

To examine the properties of the host galaxy of SN\,2020acct we perform Spectral Energy Distribution (SED) fitting, collecting archival $g$,$r$,$i$,$z$,$y$ images from the PS1 3$\pi$ catalog, alongside $J$,$H$,$K$ measurements from the 2 Micron All Sky Survey \citep[2MASS; ][]{2MASSsurvey}, W1 and W2 bands from the Wide-field Infrared Survey Explorer \citep[WISE; ][]{WISEsurvey} and NUV and FUV photometry from the Galaxy Evolution Explorer \citep[GALEX; ][]{GALEXsurvey}. We measure the global flux of the global host galaxy within a Kron-like aperture \citep{Magnier2020a} as determined within the PS1 images, applying the same aperture across all survey images and filters for consistency. We also extract the host photometry within a 1'' (corresponding to ~0.6\,kpc) aperture centered on the explosion site of the SN to determine the properties of the stellar population native to the progenitor. 

Given that SN\,2020acct is located 9.3'' from the centre of NGC\,2981, we fit both the global and local host photometry using the stellar population synthesis models of {\tt{Prospector}} \citep{Leja2017,Johnson2021}. We apply a 10\% error floor to all photometry to account for systematic uncertainties in both the photometry and the physical models being fit to the emission. For fitting the global host photometry, we invoke a model similar to that of  Prospector-$\alpha$ detailed by \cite{Leja2017}, in which the stellar mass, metallicity, six-component star formation history, and dust fraction, and reprocessing are free parameters \citep[as per][]{Ramsden2022}. In this case, the model uses a non-parametric Dirichlet star formation history (SFH) treatment, but the remaining free parameters used are well described in \cite{Leja2017}. For fitting the local environment, we assume an exponentially declining star formation history, with broad priors on stellar mass ($4<Log(M_{*})<12$), star-formation history ($-2<Log(\tau)<2$) and metallicity ($-4<Log(Z_{\odot})<10$). We generate models for both fits which include the effects of stellar and nebular emission, metallicity, dust reprocessing, sampling posterior using the Bayesian nested sampling code dynesty \citep{Speagle2020}.

Fitting the global SED, we find that NGC\,2981 has a stellar mass of $\log{(M_{\star} / M_{\odot})}=11.29^{+0.04}_{-0.06}$, and a current star formation rate of $(SFR / M_{\odot} yr^{-1}) = 17^{+15}_{-11}$. When we fit the SED of the stellar population local to the explosion site aof SN\,2020acct, within a 1'' radius of its location, we find a local
star formation rate of $\log{(SFR / M_{\odot} {\rm yr}^{-1})} = -1.1^{+0.9}_{-0.7}$ and average stellar age of $t_{\mathrm{age}}$ = 9.13$^{+4.5}_{-3.3}$ Myr, from which we infer a local specific star formation rate (sSFR) of $\log{(\mathrm{sSFR / M_{\odot} yr^{-1}})} = -10.4^{+0.8}_{-0.6}$. We show our global and local SED fits in the Appendix (Fig. \ref{extfig:SEDfits}).

We measure a local gas phase metallicity using emission lines from the LRIS spectrum taken at the location of SN\,2020acct. Only H$\alpha$, H$\beta$ and NII $\lambda$6584 lines are observed, for which we measure fluxes of $F_{H\alpha}=4.2\pm0.2\times10^{-16}$erg\,s$^{-1}$\,cm$^{-2}$,  $F_{H\beta}=7.5\pm0.8\times10^{-17}$erg\,s$^{-1}$\,cm$^{-2}$ and $F_{NII}=6.6\pm1.4\times10^{-17}$ erg s$^{-1}$\,cm$^{-2}$. We use the N2 metallicity diagnostic of $12+\log(O/H)$=8.734 + 0.462$\times\log([NII/H\alpha])$ \citep{Marino2013} and find $12+\log(O/H)$= 8.36 (0.46$Z_{\odot}$). We also estimate the local SFR using the H$\alpha$ line flux diagnostic of \citep{Kennicutt1989}, and find SFR(H$\alpha)=9.4\times10^{-3}$\sm\,yr$^{-1}$. We note that this metallicity ($\approx0.4\,Z_{\odot}$), approximate to the metallicity of the Small Magellanic Cloud, is significantly lower than the SFR inferred through our local Prospector fitting, but is likely due to differences in aperture-size. Spatially resolved integral-field spectroscopic observations will likely disentangle this issue. 

\subsection{Spectroscopic Properties}\label{sec:spec}

\subsubsection{First Peak}
From Figure \ref{fig:specev} we can see that the spectroscopic properties of the two peaks are very distinct. The higher-resolution LRIS spectrum of the first peak, obtained at $-$58\,d (approximately $-$1\,d from maximum of the first peak), displays a blue continuum with several superimposed narrow absorption and emission lines, including strong \ion{He}{1} $\lambda\lambda4471,5857$\,{\AA} and \ion{He}{2} $\lambda 4686$\,{\AA} P-Cygni features. Given the lack of corresponding H$\alpha$ and H$\beta$ emission, we identify the P-Cygni feature at $\lambda 4100$\,{\AA}  as \ion{He}{2}. 

The feature at 4600~\AA\ requires additional examination.  This feature has two peaks above the continuum with the redder peak corresponding to \ion{He}{2} $\lambda 4686$\,{\AA}.  The bluer peak is roughly consistent in wavelength with the \ion{N}{3} triplet $\lambda\lambda 4634,4640,4641$, however, we disfavor this identification, as discussed below.

The wings of each peak are much narrower than the Lorentzian (or ``IIn-like'') profiles typically seen in either flash-ionized SNe~II or SNe~Ibn \citep[which are produced by electron-scattering of free electrons as they pass through the ionized CSM;][]{Chugai2001,Huang2018}. If the two peaks were attributed to resolved narrow lines of \ion{He}{2} and \ion{N}{3}, this would imply a width of $\lesssim$1000~km\,s$^{-1}$ (see Appendix Figure \ref{fig:heiiniii}) and therefore a low optical depth for the line-forming region of the CSM compared to other CCSN progenitors \citep{Jacobson-Galan2024}. 

\begin{figure*}
\includegraphics[width=\textwidth]{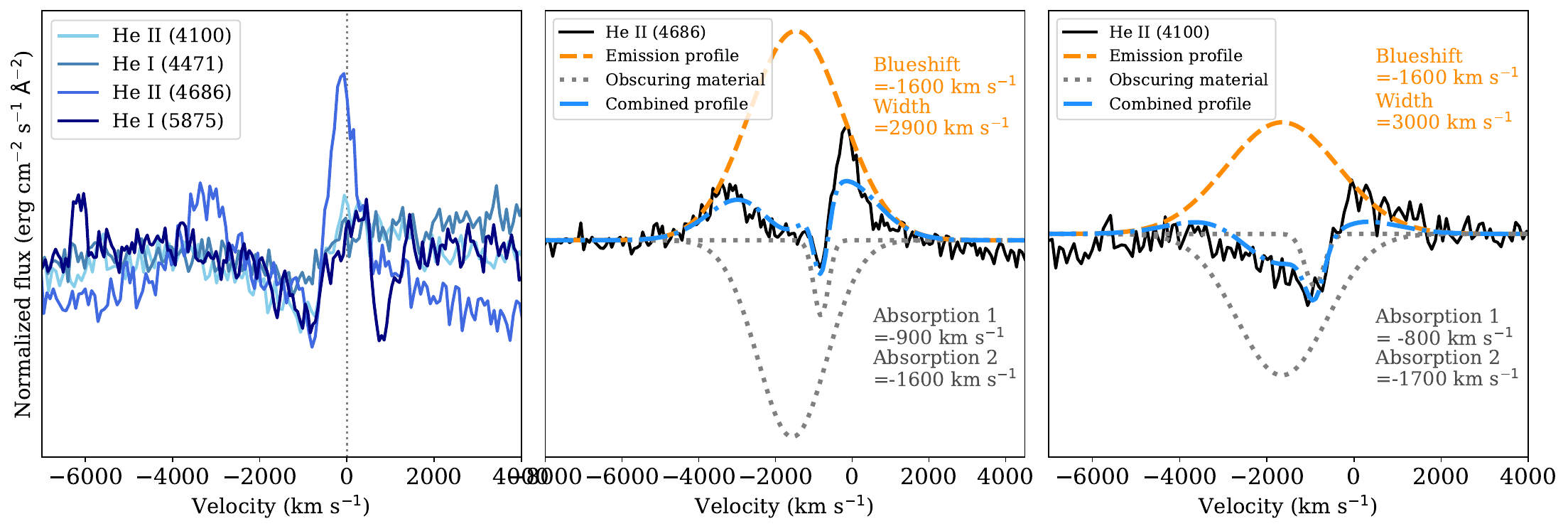}
\caption{Helium profiles in the first peak of SN\,2020acct. {\textit{Left:}} Strong asymmetric absorption features are present in all helium lines, which align in velocity space. The $\lambda$4471 line is affected by the SN continuum. {\textit{Middle:}} Fitting the \ion{He}{2} $\lambda$4868 line with a single Gaussian profile, with a width of $\sim$3000\,km\,s$^{-1}$. Two absorbing components are required, with velocities of $\sim$-900\,km\,s$^{-1}$ and $\sim$-1600\,km\,s$^{-1}$. {\textit{Right:}} The same fit to the \ion{He}{2} $\lambda4100$ line.}
 \label{fig:missinghe}
\end{figure*}

Instead, the 4600{\AA} feature can be explained exclusively as \ion{He}{2} with a complex line profile. This profile requires a single emission component blueshifted by -1600\,km\,s$^{-1}$ with a FWHM of 2900\,km\,s$^{-1}$ and two absorption components at $-900$\,km\,s$^{-1}$ and $-1600$\,km\,s$^{-1}$.  The absorption significantly reduces the core of the emission, separating it into two apparent emission peaks. This is supported by comparison to other \ion{He}{2} lines, which exhibit a consistent velocity structure (see for example  the \ion{He}{2} $\lambda 4100$\,{\AA} feature, Fig. \ref{fig:missinghe}).  The apparent blueshift of these lines from the rest frame ($\sim$1600\,km\,s$^{-1}$) may be due to destructive opacity effects \citep{HoS_Jerkstrand}, where redwardly scattered photons that are moving toward the observer are suppressed by optically thick interceding material, creating an apparent strong reduction in flux from the red wing of the line.

We thus identify the 4600\,{\AA} feature as being produced solely by \ion{He}{2} $\lambda 4686$\,{\AA} with no \ion{N}{3} contribution.  Thus the first optical spectrum of SN\,2020acct exhibits only He spectral features. These emission features have mostly disappeared by the $-$54\,d spectrum (+3\,d from first maximum), where only broad, weak \ion{He}{1} $\lambda5876$\,{\AA} emission is observed.

\begin{figure*}
\centering%
\includegraphics[width=\textwidth,trim= 0 0 0 0,clip]{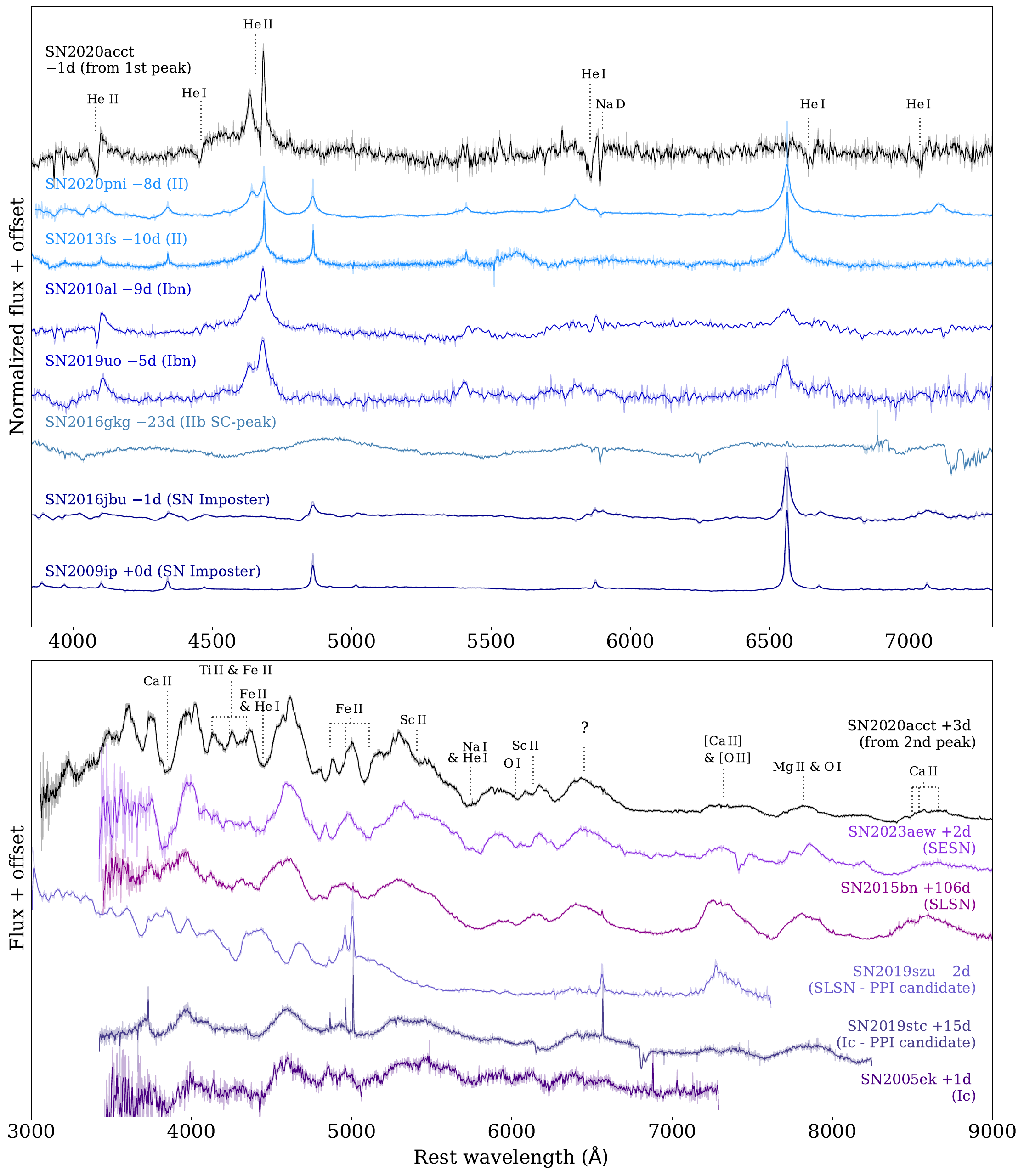}
\caption{{\textit{Top:}}  Spectroscopic comparison of the -58\,d spectrum of SN\,2020acct with `flash-ionization' spectra of young type II CCSNe \citep{Yaron2017,Terreran2022}, early spectra of SNe~Ibn \citep{Pastorello2015,Gangopadhyay2020}, SN impostors \citep{Fraser2013,Brennan2022}, and the shock cooling phase of a SNe~IIb \citep{Tartaglia2017}.  Key spectral features are marked. 
{\textit{Bottom:}} Spectroscopic comparison of SN\,2020acct around its second maximum (+3\,d) against double peaked SESN; SN\,2023aew \citep[here phase of the SN is set with respect to its second peak;][]{Kangas2024,Sharma2024}, SN\,2019stc \citep{Gomez2021}, and SN\,2019szu \citep{Aamer2023}, the superluminous SN (SLSN) SN\,2015bn \citep{Nicholl2016b}, and the fast evolving SNIc SN\,2005ek \citep{Drout2013}.}
\label{fig:speccomp}
\end{figure*}

In Figure \ref{fig:speccomp} we compare the first peak spectrum of SN\,2020acct to the those of CCSNe at similarly early epochs ($<-7$\,d). At these phases, nearby CSM is `flash-ionized' by ultraviolet (UV) radiation emitted by the breakout of the SN shock from the stellar envelope \citep{Yaron2017,Terreran2022}, rapidly cooling and recombining \citep{Gal-Yam2014,Khazov2016}. We also compare with the early flash-ionization spectra of rapidly evolving SNe~Ibn \citep{Pastorello2015,Gangopadhyay2020}, non-terminal outbursts from SN impostors \citep{Fraser2013, Brennan2022}, and with spectra of cooling of shock-heated ejecta following shock breakout (shock cooling; SC) in a SN~IIb \citep{Tartaglia2017}.

The helium emission features present within the first peak of SN\,2020acct do not resemble the broad features produced by rapidly expanding ejecta within SNe, or within SN~IIb shock-cooling peaks. The lack of Balmer emission disfavors a SN impostor interpretation for the first peak (also see Sect. \ref{sec:phot}). These features are instead more similar to those of flash-ionized CCSNe and SNe~Icn/Ibn, whose features are the result of brief CSM interaction. Though the brevity of the emission features seen in the $-$58\,d spectra (gone 4 days later) and the presence of high-ionization recombination lines such as \ion{He}{2} $\lambda 4846$\,{\AA}, are reminiscent of short-lived recombination lines in the early epochs of SNe~II \citep{Khazov2016,Bruch2023} and some SNe~Ibn \citep{Pastorello2015,Hosseinzadeh2017,Gangopadhyay2020}. The lack of H$\alpha$ emission and broader line widths (see Fig. \ref{fig:missinghe}) present in SN\,2020acct suggest a different origin of interaction. 

The broad, complex helium line profiles with emission and multi-absorption components, indicate a complex CSM structure. From our simple modeling, the absorption profiles imply that at least two regions of obscuring material sit between the emission zone and the observer in order to mask the center of the \ion{He}{2} emission profiles. The lack of corresponding hydrogen absorption suggests that this obscuring CSM is also hydrogen-free. We estimate velocities of approximately $\sim$-900\,km\,s$^{-1}$ and $\sim$-1600\,km\,s$^{-1}$ for these helium absorbing regions. The different widths and velocities of these profiles suggest a complicated geometry with multiple obscuring regions. Understanding the geometry of this CSM structure requires detailed 3D radiative transfer simulations, which is beyond the scope of this work.

\subsubsection{Second Peak}
Upon re-brightening, the second peak of SN\,2020acct exhibits a very different spectral evolution. The strong, broad absorption features visible are characteristic of a stripped envelope supernova (SESN). The lines present at this phase include: \ion{He}{1}  $\lambda$4471,4922,5876,6678\,{\AA}, \ion{Na}{1} $\lambda$5890,5896\,{\AA}, \ion{O}{1}  $\lambda$6158,7774\,{\AA}, \ion{Sc}{2}  $\lambda$5663,6247\,{\AA}, \ion{Ca}{2}  $\lambda$3934\,{\AA} and the near infrared (NIR) triplet  $\lambda$8498,8542,8662\,{\AA}, \ion{Mg}{2}  $\lambda$7887\,{\AA}, forests of blended \ion{Ti}{2}  around $\lambda$ 4400\,{\AA}, and \ion{Fe}{2}  lines around $\lambda$4550\,{\AA}. We also tentatively identify the broad, flat-topped feature around 7300\,{\AA} (obscured by a telluric in the +0\,d spectrum but clear at +3\,d) as a blend of forbidden  [\ion{Ca}{2}] $\lambda$7291,7323\,{\AA} and [\ion{O}{2}] $ \lambda$7320,7330\,{\AA}, with potential contributions from [\ion{Fe}{2}] $ \lambda$7452\,{\AA} or \ion{N}{1} $ \lambda$7468\,{\AA} toward the red wing of the feature, with the two wings becoming more distinguished by the +9d spectrum. Notably, SN\,2020acct exhibits relatively narrow absorption features. Fitting the \ion{O}{1}  $\lambda$6158\,{\AA} and \ion{Sc}{2}  $\lambda$6247\,{\AA} features (which we judge to be unblended features) in the +1 and +3\,d spectra, we find ejecta expansion velocities of $5800\pm900$\,km\,s$^{-1}$, decreasing to $4300\pm400$\,km\,s$^{-1}$ at +9\,d. The origin of the broad emission feature at 6500\,{\AA} is unclear. This could be due to H$\alpha$; however the implied velocity from the absorption component ($13,000$\,km\,s$^{-1}$) is substantially higher than the bulk of the ejecta, and we do not see any corresponding H$\beta$ emission. Other possibilities include \ion{C}{2} $\lambda$\,6580\,{\AA}, [\ion{O}{1}] $\lambda\lambda$6300,6364\,{\AA}, [\ion{N}{2}] \citep[as suggested for SN\,2023aew; ][]{Kangas2024}, or some combination of these lines. By +56\,d in the late photospheric phase, we see only rest frame emission from \ion{Na}{1}, the \ion{Ca}{2}  NIR triplet, and the ambiguous 6500\,{\AA} feature. 

In Figure \ref{fig:speccomp} we compare the +3\,d spectrum to the spectra of other double-peaked SESNe; SN\,2023aew \citep{Kangas2024,Sharma2024}, SN\,2019stc \citep{Gomez2021}, and SN\,2019szu \citep{Aamer2023}, to a late-time spectrum of the superluminous SN (SLSN) SN\,2015bn \citep{Nicholl2016b}, and to the fast evolving SN~Ic SN\,2005ek \citep{Drout2013}. The features present during the second peak of SN\,2020acct are strikingly similar to those of SN\,2023aew, and to SN\,2015bn, although less broad. The potential presence of forbidden [\ion{Ca}{2}] and [\ion{O}{2}] at maximum light is unusual for a typical SESN. Forbidden emission lines can only be produced within low-density environments, where collisional de-excitation is unlikely, and thus radiative de-excitation becomes the dominant process. As such, these features typically appear closer to the nebular phase ($>$100 d from explosion), when the ejecta become largely optically thin. SN\,2020acct joins a handful of SNe; SN\,2023aew, SN\,2019szu, SN\,2019stc and SN\,2018ibb \citep{Schulze2023}, where these features are prominent close to maximum light.  

\cite{Aamer2023} explain the early presence of [\ion{O}{2}] emission lines in SN\,2019szu as the interaction of the SN ejecta with nearby CSM, ejected 120\,d before explosion as a result of PPI, a model which has also been invoked to explain the double-peaked light curve of SN\,2019stc. We explore this further in Section \ref{sec:PPISN}. 

\subsection{Photometric Properties}\label{sec:phot}

The two peaks of SN\,2020acct are distinct from each other, with emission dropping by 2.8 magnitudes in the $r$-band within 22 days from the first maximum. In the bottom panel of Figure \ref{fig:lc} we compare the photometric evolution of SN\,2020acct to other double-peaked SESNe \citep{Gutierrez2021,Kuncarayakti2023,Das2023,Kangas2024,Sharma2024} including three PPISNe candidates \citep[SN\,2016eit, SN\,2019stc, SN\,2019szu;][]{Gomez2019,Gomez2021,Aamer2023}. We also compare the first peak's evolution to the SN~Ibn template of \cite{Hosseinzadeh2017} and to a SN impostor template \citep[constructed using data from][]{Fraser2015,Elias-Rosa2016,Tartaglia2016,Pastorello2018,Brennan2022}. 

SN\,2020acct does not closely resemble any current event within the literature. The 58\,d lag between the two peaks is similar to the peak separations in other events, however the clear drop in flux between the peaks is unique. While the second peak of SN\,2020acct falls within the typical distribution of normal Ib/Ic SN luminosities \citep{Richardson2014}, it lies toward the faint end of the double peaked SN range. The brighter first peak exceeds the luminosities typical of a SN impostor, and its light curve evolution is more rapid. Although its rapid behavior is closer to that of SNe~Ibn \citep{Hosseinzadeh2017}, the differences in spectral properties outlined in Sect. \ref{sec:spec} \citep[although see][for spectroscopic diversity]{Pursiainen2023}, and the appearance of a second peak suggests a different origin.

We construct the bolometric light curve of SN\,2020acct using the optical foreground extinction-corrected photometry (we exclude $\textit{Swift}$ photometry due to uncertain background subtraction), grouping data at 2-day intervals and fitting a simple blackbody function to the resulting light curves and integrating across the full blackbody curve to find the bolometric luminosity. We find the luminosity of the first peak to be 9.6$\pm0.6\times10^{42}$\,erg\,s$^{-1}$, and that of the second peak to be 3.52$\pm0.09\times10^{42}$\,erg\,s$^{-1}$. The flux drops to $<2\times10^{41}$\,erg\,s$^{-1}$ during the minimum between the two peaks.

Examining the blackbody fits (see Figure \ref{extfig:BB_fit}), we find that the first peak of SN\,2020acct shows very rapid evolution in both temperature and radius, cooling quickly after peak (cooling by $\gtrsim$5000\,K in 10\,days), with a radius increase of a factor $\sim$2 over 10 days to $\gtrsim8\times10^{14}$ cm. We also note the unusually large photospheric radius found for both events ($\gtrsim8\times10^{14}$\,cm), and seemingly larger for the first peak compared to the second peak. The general behavior is similar to flash-ionized SNe~II \citep[though CCSN radii are typically not so extended;][]{Irani2023}, and to Ibn/Icn SNe \citep[e.g.][]{Pellegrino2022,Pursiainen2023}, where a CSM envelope is quickly shock-heated and then cools. 

\section{Progenitor Scenarios}

We next explore different potential progenitor scenarios to see if they can reproduce the double-peaked light curve of SN\,2020acct. We first consider the plausibility of SN\,2020acct being the chance occurrence of two individual SNe originating from either the same star-forming region or binary system before exploring models from a single progenitor system. 

\subsection{Two SNe from the Same Region} \label{sec:2SNe}

We consider a scenario in which the two peaks of SN\,2020acct originate from two individual progenitor stars in the same star-forming region, which would appear to be coincident given the localization within the host galaxy. 

As the lifetimes of CCSN progenitors are short ($\sim$few Myrs), the CCSNe rate should closely follows that of current star formation rate, which we can use to estimate the likelihood of this case. We take the theoretical relationship between SFR and CCSN rate from \cite{Botticella2012}. For the local SFR within $\sim1$\,kpc of the explosion site of SN\,2020acct, which is $\sim1\times10^{-2}$\,\sm\,yr$^{-1}$, we would anticipate of order $\sim10^{-4}$\,SNe\,yr$^{-1}$. The probability of two individual SNe occurring within this star-forming region within a 60-day window is thus $\sim2.7\time10^{-10}$. The extremely low probability of two SNe occurring within 1 kpc of each other within such a short window suggests this is not a likely scenario to produce SN\,2020acct.

\subsection{Two SNe from a Binary System} \label{sec:binary}

Another possible route to producing two individual SNe at the same explosion site within a short temporal window is to have two progenitors co-evolving within a binary system. Close binary systems, in which Roche-lobe overflow occurs, have been frequently invoked as the progenitors of SESNe, as it presents a more efficient method (compared to stellar winds in single Wolf-Rayet stars) of partially stripping $10-18$\,\sm\ SN progenitors of their hydrogen envelopes \citep[e.g.][]{Claeys2011,Yoon2017}. Binaries have also been suggested as a mechanism to strip Ibn progenitor envelopes, with some potential direct progenitor detections to support this \citep{Maund2016,Sun2020}.

However, for a binary system to reproduce the observed properties of both peaks of SN\,2020acct, with both components exploding within a $\sim$60 day window, both stars within the system must evolve on similar timescales, and undergo sufficient stripping to reduce their outer hydrogen envelopes. A close system in which the binary mass ratio is high may result in unstable mass transfer while both stars are on the main sequence, leading to the formation of a common envelope with a double core \citep{Vigna-Gomez2018}. Within such a system, both stars could co-evolve on more similar timescales while efficiently removing most of their hydrogen envelopes. However, the likelihood of both components reaching the point of final core collapse within 50 days of each other seems implausible. 

Another issue with this model is the requirement that enough energy must be lost from the binary system due to its co-rotation within the envelope to sufficiently eject it before both components explode, as we do not observe signatures of strong interaction with a hydrogen-rich CSM within either peak. At the lower end of massive binaries (two 8\sm\ stars), the binding energy of the envelope is of order several $\times10^{49}$ ergs \citep{Kruckow2016}. It may be possible that the explosion of the first component releases sufficient energy to remove any envelope prior to the second explosion. However, from simple modeling ({\S\ref{sec:MOSFiTmodel}}), the kinetic energy released during the first explosion is $\sim(6-8)\times10^{48}$\,ergs, therefore unlikely to be sufficient to eject an envelope alone. It may be possible that some envelope remains after the first explosion, and any signatures of interaction are too weak to be detected while the second SN is bright. 

The extreme fine-tuning this scenario would require to evolve both binary components on similar timescales, for the common envelope to be cleanly removed before or by the first SN, and for both cores to collapse within a two-month window, makes this scenario inconceivable. Thus we do not consider this futher as a route to producing the two peaks of SN\,2020acct. 

\subsection{Rate of SN\,2020acct-like Events} \label{sec:discrate}

To help constrain potential progenitor scenarios for SN\,2020acct, we determine the observed rate of comparable double-peaked SNe to SN\,2020acct. The SN rate is defined as:

\begin{equation}
R_{\rm 20acct-like}=\frac{N}{f_{\rm spec}\,\epsilon\,VT}
\end{equation}

where $N$ is the number of events observed within a volume $V$ during the time period $T$. The efficiency of detection (the likelihood that an event is detected within the data of a given survey) is given by $\epsilon$, while $f_{\rm spec}$ is the spectroscopic completeness of the survey within $V$. 

We estimate the rate of SN\,2020acct-like events within ATLAS (our shallowest survey), over a 4-year baseline from 2018-2021, out to a distance of 300 Mpc (z$\sim$0.07), beyond which the second peak of SN\,2020acct would not have been recoverable within the ATLAS data (and the first peak would be observed as a stand-alone SN~Ibn). 

Using the ATLAS efficiency simulation tool (which accounts for the survey cadence and 5 sigma limiting magnitudes within a given field), we estimate the recovery efficiency for a SN\,2020acct template light curve \citep[see][ for full details]{Srivastav2022}. Assuming an event is efficiently `detected' if a minimum of 10 epochs of co-spatial 5$\sigma$ detections are recovered, we simulate the efficiency of ATLAS recovery by injecting 10000 events in 50 Mpc bins out to 300 Mpc. From our simulations, we find $\epsilon=0.163$ at 300 Mpc.

To estimate the spectroscopic completeness of ATLAS, we take a similar approach to \cite{Nicholl2023}, by selecting all transients detected within the survey of brightness m$<20.0$ (in either the cyan or orange filter) within a 4-year window (2018-2021) and measuring $f_{\rm spec}$ from the ratio of spectroscopically classified to total events with $>11$ ATLAS detections in the light curve (the minimum number of detections required to identify SN\,2020acct as a double-peaked event assuming an average ATLAS cadence of 1.5 days). We find $f_{\rm spec}=0.54$. 

Given the changing spectroscopic nature of SN\,2020acct, we also search for potentially missed spectroscopically classified events within the 2018-2021 ATLAS data, requiring that candidates have two peaks separated by $>$20 days (eliminating IIb shock-cooling curves), and peak luminosities within $\sim1$\,mag of each other, where any spectroscopic classification of either peak matches that of SN\,2020acct (signatures of interaction and/or a SESN). We find no other double-peaked events of a similar nature to SN\,2020acct. 

We measure the rate of double-peaked events like SN\,2020acct to be $\sim$3.3$\times10^{-8}$ events Mpc$^{-3}$\,yr$^{-1}$, approximately $\times10^{-4}$ of the CCSN rate \citep{Perley2020,Frohmaier2021}. We thus proceed under the assumption that SN\,2020acct is the product of an intrinsically rare progenitor system or explosion mechanism.

\subsubsection{Shock Cooling Modelling} \label{sec:SCmodel}

A small subset \citep[3-9\%][]{Das2023} of SNe Ib/c have pre-explosion peaks due to the shock-cooling (SC) of extended circumstellar material around the progenitor star heated by the SN shockwave following the initial explosion. We thus model the first peak of SN\,2020acct with a SC model, where photons deposited by the shockwave begin to escape from optically thick ejecta or CSM \citep{Waxman&Katz2017}. SC emission is expected to last on the order of a few days, with a light curve morphology driven by both the energetics of the shock wave and the properties of the nearby CSM \citep{Nakar&Sari2010,Waxman&Katz2017}. We fit the first peak with the revised models of \cite{Piro2021}, fitting only optical ($g,r,z,o$) photometry from the first 10-days after our estimated $t_{0}$, where the model is valid. We find the best fit to these data to be shock cooling of a CSM envelope at a radius of $R_{\rm CSM}$=750$\pm$10\,R$_{\odot}$, that has a mass of $M_{\rm CSM}$=0.20$\pm$0.05\sm, and is moving at a velocity of $v_{e}$=12,000$\pm$800\,km\,s$^{-1}$. 

The primary issue with this model is that SC emission does not provide a natural explanation for the delay between the first and second peak, and requires that the SN occur at the beginning of the first peak. It is then difficult to explain the second peak reaching maximum $\sim$50 days later. 

\subsubsection{Low Mass Helium Star Models} \label{sec:Hestarmodel}

Given the presence of helium emission during the first peak, we explore models commonly used to describe the production of SNe~Ibn, an intrinsically rare subset of SESN \citep[approximately 1\% of all SNe;][]{Pastorello2008}. A massive Wolf-Rayet star which loses all or most of its hydrogen envelope through luminous blue variable (``LBV-like'') outbursts \citep[e.g.][]{Pastorello2007}, winds, or binary interaction can accumulate a substantial helium-rich CSM around the progenitor. Interaction of the final SN ejecta with this CSM \citep[e.g.][]{Foley2007,Pastorello2008,Dessart2022} could potentially power the light curve.

We explore several models involving explosive mass ejection and winds from low mass (initial masses 2-3\,M$_{\odot}$) helium stars at solar metallicity \citep[see also][]{Dessart2022}. Two examples are shown in the Appendix (Figure \ref{extfig:helowmass}). One, based on a helium star with initial mass 2.6\,M$_{\odot}$ (pre-SN mass 2.14\,M$_{\odot}$) experiences a very low energy terminal explosion ($2.3\times10^{49}$\,erg), that interacts with a shell ejected by a silicon flash 101 days earlier. The other, a helium star with initial mass 2.7\,M$_{\odot}$ (pre-SN mass 2.21\,M$_{\odot}$) experiences no silicon flash, but instead interacts with a very strong wind (0.1\,M$_{\odot}$\,yr$^{-1}$). Such a large mass loss rate might be reasonable for a star that is burning oxygen and silicon during the last year of its life in unstable flames bounded by a molecular weight inversion \citep{Woosley2019}, or perhaps late-stage binary interaction.

While both models are capable of reproducing the first peak of SN2020acct, each has some deficiencies. Models invoking interaction of a terminal explosion with a previously ejected shell require very low explosion energies ($2.3\times10^{49}$\,erg) to avoid making the light
curve too bright. Reducing the wait time between the silicon flash and core collapse could produce a fainter event, but then the SN ejecta would also overtake the ejected shell before peak
luminosity was reached, resulting in no narrow emission lines. Moreover, this model does not
offer a natural explanation for the presence of the second peak. 

The simplest wind model produces a light curve that declines too slowly following the first peak. This could reflect the decreasing optical efficiency of the forward shock as it moves outside the photosphere or a variable mass loss rate. In particular, if the mass loss rate increased with time just before the explosion, the density in the wind would decline more steeply than $r^{-2}$. This model also does not provide an intuitive explanation for the second peak. While a composite model invoking an interaction with a shell of material ejected by a silicon flash prior to the wind phase is possible, the lack of narrow lines during the second peak discredits this. 

While these models arise from a more instinctive progenitor population given the host environment of SN\,2020acct (see Sect. \ref{sec:host}), they critically cannot explain the presence and properties of the second peak. We thus do not consider these models further here.

\subsubsection{MOSFiT Modelling} \label{sec:MOSFiTmodel}

We use the Modular Open Source Fitter for Transients ({\tt{MOSFiT}}) code \citep{Guillochon2018} to estimate the likely explosion parameters for each peak under the assumption that the first peak is CSI-powered \citep{Chatzopolos2012,Jiang2020} and the second is the result of Ni$^{56}$ decay model \citep{Arnett1982}. We select data within the MJD range 59192.65 - 59223.43 for the first peak and in the range 59248.40 - 59280.27 for the second peak, independently fitting the two models but allowing for the explosion epoch of the second to coincide with the first peak. For the CSI modeling, we explore different CSM density profiles ($\rho_{\mathrm{CSM}}$), where the density is a power law function of the radius, $r$, such that $\rho_{\mathrm{CSM}} \propto r^{-s}$. We fit the peak three times under the assumption of (i) a shell-like CSM ($s=0$), (ii) a wind-like CSM ($S=2$), and (iii) with $s$ as a free parameter. To efficiently sample our posterior parameter space, we use the Bayesian nested sampling code dynesty \citep{Speagle2020}. 

Overall we find that the CSI and the Ni$^{56}$ decay models provide agreeable fits to the data. In particular, the CSM interaction model is able to capture the swift rise and slower decline of the first peak, although with a very low CSM mass of 0.05\,M$_{\odot}$ and 7$\times10^{-3}$\,M$_{\odot}$ of ejecta. We find negligible differences in the goodness of fit when using a wind-like and a shell-like CSM, and leaving this parameter free does not suggest a preference either way. However, as the CSI models implemented within {\tt{MOSFiT}} are based on the analytic models of \cite{Chatzopolos2012}, which assume a continuous, power-law density profile for both ejecta and CSM, these unusual values may be a result of trying to fit a simple model to a more complex physical scenario.  For the second peak, we find good fits to the Ni$^{56}$ decay models involving 0.1\,M$_{\odot}$ of ejecta with an extremely high nickel fraction (f$_{\rm Ni}=0.91$). Such low ejecta mass is unusual for a SESN. SNe~Ibc typically eject between 2-4\,M$_{\odot}$ of material \citep{Drout2011,Lyman2016}, of which  0.08\,M$_{\odot}$ on average is Ni$^{56}$ \citep{Afsariardchi2021}. Though Arnett Ni$^{56}$ masses are usually overestimated \citep{Afsariardchi2021}, the low ejecta mass and unrealistically high Ni$^{56}$ could indicate that the second peak is not produced by a standard SESN progenitor. Our assumed model priors and resulting fit parameters can be found in Appendix Table \ref{tab:mosfit} and our fit shown in Appendix Figure \ref{fig:mosfit}. 

\subsection{Discussion}
Based on its spectroscopic properties exhibited by both peaks and the results of our {\tt{MOSFiT}} modeling, it is reasonable to assume that the first peak of SN\,2020acct is powered solely by CSM interaction. Additionally, the {\tt{MOSFiT}} results imply that the second peak explosion date occurs {\emph{after}} the first peak has faded, and its spectral evolution is similar to that of a SESN powered by Ni$^{56}$ decay (although the unrealistically high nickel fraction and low nickel mass make this scenario unlikely, see Section 5 for further discussion).
 
However, both peaks of SN\,2020acct present distinct properties that make their origins difficult to reconcile with normal SN models. The second peak of SN\,2020acct appears to be a terminal explosion, both from its spectroscopic properties and from its photometric and spectroscopic similarity to SESNe. Despite its spectroscopic similarity to SESNe and alongside light curve properties concurrent with the other events within this class, the relatively low expansion velocities of the ejecta, alongside the early emergence of forbidden [\ion{Ca}{2}] and [\ion{O}{2}] lines typically observed at the nebular phase, suggest a different progenitor origin to a typical SESN. 

The first peak is equally problematic. Its spectroscopic features suggest strong interaction with nearby hydrogen-free CSM, although its duration and luminosity make it difficult to reconcile as either a stand-alone SN or as a SN impostor. Its separation in time from the second peak (58\,d in the rest frame) puts it at odds with shock-cooling models from shock-breakout at the onset of core collapse, and from models invoking CSM interaction from explosions of low-mass helium stars. In particular, if originating from the same explosion, the separation between the two peaks is seemingly too long for them to be causally connected (i.e. the onset of core collapse cannot happen long before the start of the second peak). The strong, broad helium emission lines imply a different CSM configuration to the nearby CSM illuminated during the early phases of normal CCSNe and SNe~Ibn. 

PPISN is a very appealing scenario to describe SN\,2020acct, as its properties align with the key hallmark of PPISNe - the presence of an interaction-powered transient with SN-like energies prior to a terminal explosion. PPI would provide a natural explanation for the two luminous and distinct peaks of the event, whilst the low observed rate of similar events measured within ATLAS would agree with the natural rarity of PPI progenitors \citep[estimated to be a few in a thousand of all CCSN progenitors][]{Vigna-Gomez2019}. We explore a PPI progenitor scenario for SN\,2020acct more in the following section.

\section{Pulsational Pair Instability} \label{sec:PPImodel}

Episodic eruptive mass-loss in massive stars has been shown to produce luminous precursory peaks within SN light curves \citep{Smith2010a,Fraser2013,Smith2014,Bilinski2015,Elias-Rosa2016}. However, the luminosity (peaking on average at $\sim10^{42}$erg\,s$^{-1}$), evolutionary timescale and distinct spectroscopic features of the first peak make it unlikely that the first peak arises due to an LBV-like eruption as seen in SN impostor events. The absence of hydrogen in the first peak spectra of SN\,2020acct also makes it difficult to explain as an LBV eruption.

PPI presents one possible way of producing a peak powered purely by helium-rich CSM interaction without needing to invoke an underlying SN. Collisions of ejected shells of material are capable of powering transients ranging between $10^{41} - 10^{44}$\,erg\,s$^{-1}$ \citep{Woosley2017,Renzo2020}, prior to the onset of final core collapse. However, due to the requirement of large initial masses to build a sufficiently massive core, stars which experience PPI are expected to be intrinsically rare, thus the rate of PPISNe should be low \citep{Renzo2020}. 

In this section, we compare the light curve of SN\,2020acct to PPI models of single helium cores and explore the feasibility of this progenitor scenario with regard to its observed properties and its host environment.

\subsection{PPI Models}

Theoretical models suggest that PPI is capable of reproducing a wide variety of light curve morphologies and spectra \citep{HegerWoosley2002,Moriya2015,Woosley2017,Renzo2020}. For single-star models, this observational diversity is a result of differences in the efficiency of stripping the outer hydrogen envelope, while the number of pulses is governed by the energetics of the initial pulse (such that more energetic pulses are less frequent). For stars $\gtrsim40$M$_{\odot}$, pulses from PPI are expected to completely remove the outer hydrogen envelope of the progenitor, leaving only a bare He core at the point of collapse. Early emission is expected in PPI events as a result of colliding shells of ejected material. In cores $>$44\sm, the pulse interval should become long enough ($\sim$several days) that individual peaks will be discernible in the light curve, although above $\gtrsim$52\sm, the delay between the pulses are too far apart in time to produce bright collisions with the first ejected material \citep{Woosley2017}. 

For a star stripped of its hydrogen envelope, the observed duration of the double-peaked light curve (at least 60\,d), qualitatively agrees with models of helium core masses entering PPI of $\sim$50\,M$_{\odot}$ \citep[][see Fig. 5]{Woosley2017}, which we use as the basis for our modeling here. Including the additional mass ejection episodes to create the circumstellar shell that produces the narrow lines in the first peak, the upper limit on this helium core mass is about 53\,M$_{\odot}$ \citep[for the nuclear reaction rates assumed in][]{Woosley2017}. These core masses correspond to larger helium star masses when the core is first uncovered, either via winds or binary interaction, with the actual values dependent on metallicity and an uncertain mass loss rate. Assuming a metallicity 10\% solar and the WR mass loss rates of \cite{Yoon2017}, the initial helium core mass would be $70 - 73$\,M$_{\odot}$, corresponding to a Zero Age Main Sequence mass of $\sim140 - 150$\,M$_{\odot}$ \citep{Woosley2019}. In this mass range, the PPI models transition from those which experience multiple pulses (50\,M$_{\odot}$ models), each with a peak luminosity $\sim$10$^{43}$\,erg\,s$^{-1}$, to just a single major outburst (cores $>$53\,M$_{\odot}$). In more massive cores the initial pulses are stronger, resulting in a longer delay until the final death of the star, but also burning more of the carbon and oxygen within the core. As a result, the final pulses are fewer and weaker. An uncertain, but diminishing mass of carbon also acts to decelerate the collapse, weakening the pulses\footnote{Uncertainties in convection \citep{Renzo2020} could have a similar, or even larger, dampening effect.}.

\begin{figure*}
\centering
\includegraphics[width=0.8\textwidth,trim=2.cm 1.5cm 2.cm 2.5cm,clip]{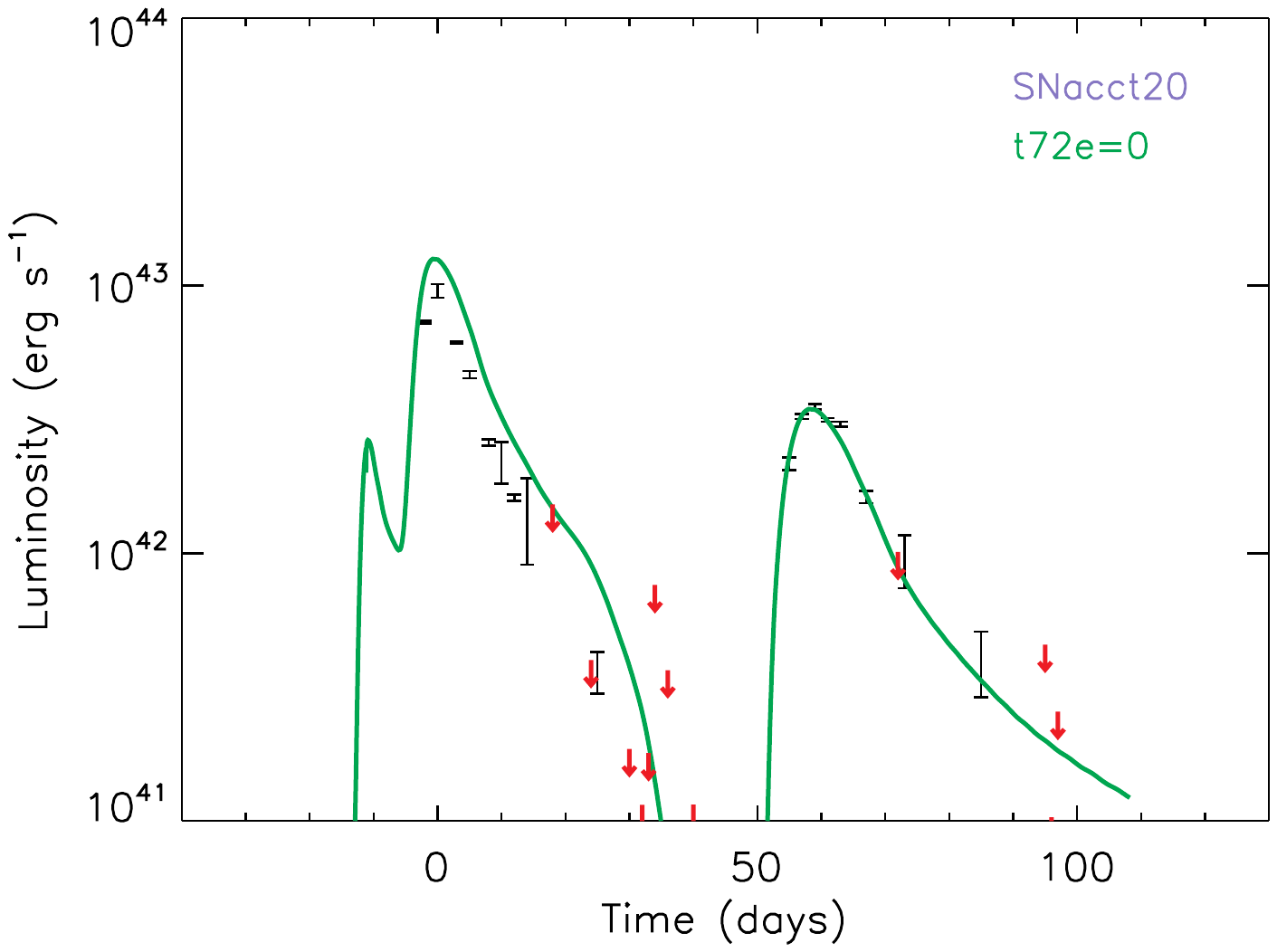}
\caption{Pulsational pair instability modelling of SN\,2020acct, compared to its bolometric light curve. From a 72M$_{\odot}$ He core (corresponding to a mass at the onset of PPI of 52.9\,M$_{\odot}$) the first peak is the result of colliding shells produced by three quick successive pulsations, while the second peak is produced by a weak final pulse which induces final collapse of the core to a 47.5\,M$_{\odot}$ black hole. The weak model peak before the first observed peak comes from the collision of closely spaced pulses 2 and 3, and its actual presence is uncertain.}
\label{fig:PPI}
\end{figure*}

Here we compute models of helium stars with helium core masses at the onset of PPI in the range $51-53$\,M$_{\odot}$ using the {\sc{kepler}} code \citep{Weaver1978,Woosley2017}, assuming a metallicity of 10\% solar, evolved with mass loss \citep[mass loss models as per ][]{Woosley2019}. To encapsulate the relative luminosity and separation of the two peaks of SN\,2020acct, we modify the reaction rates from \cite{Woosley2017} by a factor of 1.2 compared to the default 3$\alpha$ and $^{12}$C($\alpha,\gamma$)$^{16}$O reaction rates to increase the carbon mass fraction at central helium depletion to 0.124. This variation is allowed by uncertainties in the rates \citep{WoosleyHeger2021}, and acts to shorten the delay between the initial pulse and the death of the star. We add an additional degree of freedom by allowing the shock energy and interval between the peaks to vary by artificially varying neutrino energy losses, which act to shorten the cooling time between pulses. We demonstrate the effects of these adjustments in the Appendix (Fig. \ref{extfig:PPIhe71}). A detailed discussion of these models will be presented in Woosley et al. ({\it{in preparation.}}). We present our best fitting model (\lq he72\rq) in Figure \ref{fig:PPI}, in which pulsations of a helium core of 72\,M$_{\odot}$ (corresponding to core mass at the point of PPI of 52.92\,M$_{\odot}$) are found to recreate the two peaks of SN\,2020acct. 

In our best-fitting model, the modified core experiences 5 pulses before collapse. The first pulse, powered mostly by explosive carbon burning, reaches a maximum central temperature of 2.96$\times10^{9}$\,K, ejecting 0.90\,M$_{\odot}$ of material, with a kinetic energy 7.7$\times10^{49}$\,erg. This mass ejection is faint, as no significant radioactive material is ejected and adiabatic expansion reduced the energy deposited by the shock.  Most of the energy from nuclear burning ($\sim2\times10^{51}$\,erg), goes into expanding the star, reducing its central temperature, after some oscillations, to 8.45$\times10^{8}$\,K. At this low temperature, it requires 11 years of neutrino-mediated Kelvin-Helmholtz evolution for the core to once again become hot enough to be unstable. During this, most of the energy deposited by the first pulse is radiated away as neutrinos.

Pulses 2, 3, and 4 then occur in rapid succession, separated by only 0.46 days and 3.3 days, respectively (reaching core temperatures of 3.11, 3.31, and 3.89$\times10^{9}$\,K). Combined, these three quick pulses eject 3.90\,M$_{\odot}$ of material with a kinetic energy of 3.8$\times10^{50}$\,erg. Once again, the majority of the nuclear energy goes into the expansion of the star. It is the collision of pulse 4 with the combined ejecta of pulses 2 and 3 that produces the first peak of SN\,2020acct. After these last three pulses, the central temperature of the core is 1.23 $\times10^{9}$\,K. At this temperature, the associated waiting time until the death of the star in the unmodified model would be 110 days. Here we temporarily increase the neutrino losses\footnote{Neutrino losses scale with temperature as $\propto\,T^{14}$.} by a factor of 1.8, which shortens the waiting by 61 days (see Fig. \ref{extfig:PPIhe71}), and introduces a weak additional fifth pulse.

The fifth and final pulse reaches a temperature of 5.8 $\times10^{9}$\,K and then rebounds weakly, producing an out-going shock of only about 2$\times10^{49}$\,erg (corresponding to a luminosity of  4$\times10^{42}$\,erg\,s$^{-1}$, comparable with the second peak of SN\,2020acct). This weak pulse ejects an additional 0.6\,M$_{\odot}$ of material at low velocity, before the core finally collapses 30 minutes later, producing a black hole with baryonic mass 47.5\,M$_{\odot}$. For many other models of similar core mass, a slightly higher final bounce temperature is produced ($\sim$6$\times10^{9}$\,K), which causes the core to immediately collapse into a black hole. It is also possible that this immediate collapse followed by low-level accretion could power an additional weak outburst, but without additional assumptions, there would be no second peak in the light curve (Woosley et al. {\textit{in prep.}}). 

This model presents a good overall fit to the bolometric light curve of SN\,2020acct in Figure \ref{fig:PPI}. We note that the weak peak prior to the first major peak (luminosity $\sim$2$\times10^{42}$\,erg\,s$^{-1}$) is due to the collision of the closely spaced pulses 2 and 3, and may not be present given uncertainties in timing due to neutrino cooling. Given these additional uncertainties and parameters necessary to the \lq modified\rq\, model, we treat the he72 model as suggestive of the ability to describe SN\,2020acct as a PPISN, rather than a final solution.

\subsection{Consistency with Observed Properties} \label{sec:PPISN}

Both qualitatively and quantitatively, SN\,2020acct fits the description of PPI well. The light curve shows a strong match to PPI from a 72\,M$_{\odot}$ helium core in which a shell-shell collision is observed from material ejected in a combination of three pulses which occurred 64 days prior to the final collapse of the core. We note the large difference in the inferred CSM mass from {\tt{MOSFiT}} modeling (0.05\sm) of the first peak and the total CSM mass from shells 2, 3, and 4 (3.9\sm), whose collisions produce the first peak in our radiative transfer models. This difference is likely to arise from the inefficient conversion of kinetic energy to optical radiation within the PPI model, combined with the density structure of the CSM. If the location of the photosphere is not reflective of where the bulk of the ejected material is, analytical light curve models where the luminosity follows the forward shock may not accurately trace all of the ejected CSM for a PPISN.

The photospheric radius and effective temperature for the first peak of the PPI model are 6$\times10^{14}$\,cm and 15,000\,K, respectively. While preliminary (pending a more careful treatment of radiation transport), these values agree with the properties of SN\,2020acct (Appendix Figure \ref{extfig:BB_fit}). The interacting, fast-evolving spectroscopic features of the first peak, with strong helium emission, can be explained under shock heating of colliding shells of CSM. PPI typically predicts shell velocities of $2000-4000$\,km\,s$^{-1}$ \citep{Woosley2017}, which would agree with the widths of the unobscured profiles from our modeling in Section \ref{sec:spec}. The absorbing material in front of the photosphere could originate from material ejected from the progenitor in previous outbursts. While the inferred velocities of the absorbing material ($\sim900$\,km\,s$^{-1}$ and $\sim1600$\,km\,s$^{-1}$) are not entirely consistent with the shell velocity predictions of the PPI model. However, uncertainties in the assumptions behind the symmetry of the pulse or the efficiency of the ejection may reconcile these. 

From both PPI modeling and simple blackbody fitting, we can see that the apparent photosphere occurs at much larger radii ($6\times10^{14}$cm) than typically assumed for a CCSN \citep[$\sim10^{14}$cm;][]{Irani2023}, indicating farther removed material than that involved with normal interacting CCSNe. The unusual properties of the second peak can also be accounted for under PPI. The early presence of forbidden [\ion{Ca}{2}] and [\ion{O}{2}] emission could originate from the SN ejecta crossing the radius at which the shell collision occurred, interacting with the low-density CO-rich material ejected by earlier pulses. Forbidden line features have also been observed within the double-peaked PPISN candidates SN\,2019stc \citep{Gomez2021} and SN\,2019szu \citep{Aamer2023}, the latter in particular within its early (near maximum light) spectra, similar to SN\,2020acct.

\subsection{Environment and Progenitor}

PPI can provide an intuitive explanation for the two distinct peaks of SN\,2020acct. We now investigate whether the local environment of SN 2020acct is consistent with PPI models. As PPI requires massive helium cores for pair production (M$_{\rm He}>\sim$30\sm), we typically expect to find PPISN progenitors within low metallicity environments \citep{Yusof2013,Vink2018}, and given their expected short lifetimes ($\sim$\,Myrs), typically associated with regions of strong star formation. The local 1'' radius environment around SN\,2020acct presents average star forming properties compared to the bulk CCSN population \citep{Irani2022}, and a relatively low sSFR for a CCSN host. From the host galaxy spectrum we estimate a local metallicity \citep[as per][]{Storchi-Bergmann1994}, and find it to be 12+log(O/H)=8.36, or 0.4\,Z$_{\odot}$. While this estimate is sub-solar, it is still significantly higher than the $\sim$10\% solar metallicity assumed in most single-star PPI models \citep{Yoshida2016,Woosley2017,Renzo2020}. 

However, it has been suggested by \cite{Vigna-Gomez2019} that helium cores massive enough to reach the PPI threshold may be formed via the merger of a binary system composed of less massive stars (ZAMS between $40-64$\sm). After an extended period of hydrogen fusion in both stars cores, the two stars merge to form a single star whose newly combined helium core is sufficiently massive to be unstable to PPI, and will later undergo pulsations. \cite{Vigna-Gomez2019} invoke this model as a potential route to producing PPISN where the progenitors retain more of their hydrogen envelope, than within single-star models of the same core mass for a given metallicity. Thus binary mergers may provide a route to produce PPISN under higher metallicity conditions without needing to form an exceptionally massive single star progenitor in situ. 

To estimate the probability of producing a 72\,M$_{\odot}$ helium core within the local host environment of SN\,2020acct, we use the physical properties inferred from our PPI modeling to search for potential progenitor systems within the Binary Population and Spectral Synthesis data release (BPASSv2.2.2; \citealt{Eldridge2017,Eldridge2018,Stevance2020}). Assuming a metallicity of $Z=0.004$, ($0.4\,Z_{\odot}$) we set the hydrogen mass to M$_{\mathrm{H}}<0.01$\,M$_{\odot}$ and the hydrogen mass fraction to $X<0.001$ \citep[i.e. hydrogen-stripped cores;][]{Dessart2012}. We then search for models between 70 and 75\,M$_{\odot}$. Weighting the models by the Initial Mass Function of \cite{Kroupa2001} and binary fractions and period distributions of \citealt{Moe2017}, we find 0.7 systems per million solar masses that meet the above criteria at this metallicity. Though these stars are predicted to be within binary systems, none are the result of a merging system. 

Though we do not place any constraints upon the ejecta mass of the final SN within the models of our BPASS search, we find that the average IMF weighted ejecta mass to be $M_{\rm ej}=5.54$\,M$_{\odot}$. Pulsations are not modelled within the BPASS framework, however this ejecta mass agrees with the estimated total mass lost from pulsations (5.4\,M$_{\odot}$).

Using the double-Schechter galaxy stellar mass function of \cite{Wright2017} to determine the total mass of galaxies, and assuming that the main sequence lifetime scales with stellar mass as $t\propto M^{-2.5}$, we calculate the rate from transients originating from a 72\,M$_{\odot}$ helium core to be $1.1\times10^{-8}$ events Mpc$^{-3}$\,yr$^{-1}$. This is in agreement with the upper limit of our measured rate from ATLAS. Both our ATLAS rate and estimate from viable BPASS progenitor models are approximately an order of magnitude lower than theoretical predictions from \cite{Vigna-Gomez2019}, who estimate PPISN from binary mergers should be a few for every thousand CCSNe. Though a merged binary system provides an easier route to producing the required core mass for PPI within the host galaxy environment of SN\,2020acct, it is not impossible that the progenitor was produced within a pocket of low metallicity within the galaxy, in which a single massive (M$_{\rm ZAMS}\approx$150\,M$_{\odot}$) progenitor is more easily born. High-resolution environmental studies, in which the very local ($\sim$10's pc) environment to SN\,2020acct can be probed, are the best way to determine the feasibility of this latter scenario (Angus et al. {\textit{in prep.}}).

\section{Conclusions} \label{sec:conclude}

We have presented the photometric and spectroscopic follow-up of the unusual transient SN\,2020acct in NGC\,2981 at $z=0.034$. This event shows a distinct double-peaked light curve, with peaks separated by 58\,d in the rest frame, both astrometrically coincident within the same region of a massive host galaxy. The first peak is more luminous and rapidly evolving, exhibiting strong helium emission lines in its spectra, which disappear quickly. The second peak shows behavior similar to a SESN, but with relatively low expansion velocities and forbidden lines around maximum light. From our analysis of the data and exploration of possible progenitor scenarios, we conclude the following: 

\begin{itemize}
    \item The properties of the first peak of SN\,2020acct are consistent with a transient powered by hydrogen-poor CSM interaction. Spectroscopically it most closely resembles an interacting hydrogen-poor SNe~Ibn, although its helium emission features, partially obscured by foreground helium CSM, are much broader than typical Ibn events. Modeling the photometric properties of the peak suggests this interaction occurs at some distance ($8\times10^{14}$\,cm) from the progenitor.
    
    \item The second peak shows signatures of a terminal stellar explosion, i.e.~absorption features similar to SNe~Ibc. The presence of forbidden [\ion{Ca}{2}] and [\ion{O}{2}] lines around maximum light indicate the presence of low-density, pre-burnt material external to the explosion, being swept up with the SN ejecta. 
    
    \item We find the rate of double-peaked events like SN\,2020acct to be $\sim$3.3$\times10^{-8}$ events Mpc$^{-3}$\,yr$^{-1}$ at $z=0.07$ ($\sim0.001$\% of the CCSN rate).
    
    \item We do not find it likely that SN\,2020acct is the random occurrence of two SNe from within the same star-forming region of NGC\,2981, given the relatively low star formation rate of the local environment. 
    
    \item We rule out the possibility that SN\,2020acct originates from the rapid successive explosions of stars within a close binary system. Though binary co-evolution can evolve two stars on a similar timescale, significant tweaking would be required to reproduce the observed properties and explosion times of both peaks. 
    
    \item The interaction of a low mass $2.5 - 3.0\,M_{\odot}$ helium star with its late-stage wind and/or mass ejections is deemed an unlikely progenitor route, given the inability of these models to produce a secondary peak without signatures of interaction.
    
    \item We pose that SN\,2020acct may be a candidate PPISN, with the first peak being driven by shock collision of previously ejected shells of material. We fit the bolometric light curve of SN\,2020acct with models of helium core explosions, and find that it can be described well with pulsations of a 72\,M$_{\odot}$ core. 
    
    \item The relatively metal-rich and low-star-forming nature of the local host environment seems at odds with a PPI interpretation. BPASS models indicate that 0.7 stripped helium cores of the necessary mass are created per million solar masses of material within such environments. These progenitor numbers are in approximate agreement with the measured upper limit to the rate of SN\,2020acct-like events from ATLAS. It is unclear whether such PPI progenitors may be formed in situ at higher metallicity, or must be produced via close binaries mergers \citep[e.g.][]{Vigna-Gomez2019}, with further detailed studies of the local environment required to settle the discrepancy. 
\end{itemize}

The increased survey volume from Rubin Observatory Legacy Survey of Space and Time (LSST) will boost the transient discovery rate by an order of magnitude, while also enabling the detection of faint light curve structures such as pre-explosion outbursts fainter than that in SN 2020acct. Identification and follow-up of double-peaked SNe from Rubin will improve our understanding of the rate of PPISNe, the diversity in their properties, and the range of metallicities that can support them. Since these massive stars are likely the progenitors of at least some of the black holes detected through gravitational wave emission, follow-up of PPISN candidates will provide a powerful, orthogonal constraint on the evolutionary pathways for these systems.

\begin{acknowledgments}
C.R.A.\ thanks S.\ Srivastav for the excellent pun contributions to the title of this work.

C.R.A., M.N.\ and A.A.\ are supported by the European Research Council (ERC) under the European Union’s Horizon 2020 research and innovation programme (grant agreement No.~948381) and by UK Space Agency Grant No.~ST/Y000692/1.
The UCSC team is supported in part by NASA grant 80NSSC20K0953, NSF grant AST--1815935, STScI grant HST-SNAP-16691, the Gordon \& Betty Moore Foundation, the Heising-Simons Foundation, and by a fellowship from the David and Lucile Packard Foundation to R.J.F.
V.A.V.\ acknowledges support by the National Science Foundation through AST--2108676 and under Cooperative Agreement PHY--2019786.
M.P.\ acknowledges support from a UK Research and Innovation Fellowship (MR/T020784/1).
P.R.\ acknowledges support from STFC grant 2742655.
H.F.S.\ is supported by the Eric and Wendy Schmidt A.I.\ in Science fellowship.
M.R.S.\ is supported by the STScI postdoctoral fellowship.
A.G.\ is supported by the National Science Foundation under Cooperative Agreement PHY--2019786 (The NSF A.I.\ Institute for Artificial Intelligence and Fundamental Interactions, http://iaifi.org/).
A.C.\ acknowledges support from ANID’s Millennium Science Initiative through grant AIM23-0001.
M.R.D.\ acknowledges support from the NSERC through grant RGPIN-2019-06186, the Canada Research Chairs Program, and the Dunlap Institute at the University of Toronto.
C.G.\ and D.F.\ are supported by a VILLUM FONDEN Young Investigator Grant (project number 25501), and by research grants (VIL16599, VIL54489) from VILLUM FONDEN.
G.N.\ gratefully acknowledges NSF support from AST--2206195, and a CAREER grant AST--2239364, supported in-part by funding from Charles Simonyi, and OAC--2311355, DOE support through the Department of Physics at the University of Illinois, Urbana-Champaign (13771275), and support from the HST Guest Observer Program through HST-GO-16764. and HST-GO-17128 (PI: R.\ Foley).
S.J.S.\ acknowledges a Royal Society Fellowship and grants ST/Y001605/1.
Parts of this research were supported by the Australian Research Council Discovery Early Career Researcher Award (DECRA) through project number DE230101069. 

The Young Supernova Experiment (YSE) and its research infrastructure is supported by the European Research Council under the European Union's Horizon 2020 research and innovation programme (ERC Grant Agreement 101002652, PI K.\ Mandel), the Heising-Simons Foundation (2018-0913, PI R.\ Foley; 2018-0911, PI R.\ Margutti), NASA (NNG17PX03C, PI R.\ Foley), NSF (AST--1720756, AST--1815935, PI R.\ Foley; AST--1909796, AST-1944985, PI R.\ Margutti), the David \& Lucille Packard Foundation (PI R.\ Foley), VILLUM FONDEN (project 16599, PI J.\ Hjorth), and the Center for AstroPhysical Surveys (CAPS) at the National Center for Supercomputing Applications (NCSA) and the University of Illinois Urbana-Champaign.

Pan-STARRS is a project of the Institute for Astronomy of the University of Hawaii, and is supported by the NASA SSO Near Earth Observation Program under grants 80NSSC18K0971, NNX14AM74G, NNX12AR65G, NNX13AQ47G, NNX08AR22G, 80NSSC21K1572, and by the State of Hawaii.  The Pan-STARRS1 Surveys (PS1) and the PS1 public science archive have been made possible through contributions by the Institute for Astronomy, the University of Hawaii, the Pan-STARRS Project Office, the Max-Planck Society and its participating institutes, the Max Planck Institute for Astronomy, Heidelberg and the Max Planck Institute for Extraterrestrial Physics, Garching, The Johns Hopkins University, Durham University, the University of Edinburgh, the Queen's University Belfast, the Harvard-Smithsonian Center for Astrophysics, the Las Cumbres Observatory Global Telescope Network Incorporated, the National Central University of Taiwan, STScI, NASA under grant NNX08AR22G issued through the Planetary Science Division of the NASA Science Mission Directorate, NSF grant AST-1238877, the University of Maryland, Eotvos Lorand University (ELTE), the Los Alamos National Laboratory, and the Gordon and Betty Moore Foundation.

Parts of this research are based on observations made with the Nordic Optical Telescope (programme 62-507; PI Angus) owned in collaboration by the University of Turku and Aarhus University, and operated jointly by Aarhus University, the University of Turku and the University of Oslo, representing Denmark, Finland and Norway, the University of Iceland and Stockholm University at the Observatorio del Roque de los Muchachos, La Palma, Spain, of the Instituto de Astrofisica de Canarias.  Data presented here were obtained in part with ALFOSC, which is provided by the Instituto de Astrofisica de Andalucia (IAA) under a joint agreement with the University of Copenhagen and NOT.
A subset of the data presented herein were obtained at the W.\ M.\ Keck Observatory. NASA Keck time is administered by the NASA Exoplanet Science Institute. The Observatory was made possible by the generous financial support of the W.\ M.\ Keck Foundation. The authors wish to recognise and acknowledge the very significant cultural role and reverence that the summit of Maunakea has always had within the indigenous Hawaiian community.  We are most fortunate to have the opportunity to conduct observations from this mountain.
Research at Lick Observatory is partially supported by a generous gift from Google.
This research is based on observations made with the NASA/ESA Hubble Space Telescope obtained from the Space Telescope Science Institute, which is operated by the Association of Universities for Research in Astronomy, Inc., under NASA contract NAS 5--26555. These observations are associated with programs GO--16657 and SNAP--16691.
This work has made use of data from the Asteroid Terrestrial-impact Last Alert System (ATLAS) project. ATLAS is primarily funded to search for near earth asteroids through NASA grants NN12AR55G, 80NSSC18K0284, and 80NSSC18K1575; byproducts of the NEO search include images and catalogs from the survey area.  The ATLAS science products have been made possible through the contributions of the University of Hawaii Institute for Astronomy, the Queen's University Belfast, and the Space Telescope Science Institute.
Based on observations obtained with the Samuel Oschin Telescope 48-inch and the 60-inch Telescope at the Palomar Observatory as part of the Zwicky Transient Facility project. ZTF is supported by the National Science Foundation under Grant Nos. AST-1440341, AST-2034437, and a collaboration including Caltech, IPAC, the Weizmann Institute for Science, the Oskar Klein Center at Stockholm University, the University of Maryland, the University of Washington, Deutsches Elektronen-Synchrotron and Humboldt University, the TANGO Consortium of Taiwan, the University of Wisconsin at Milwaukee, Trinity College Dublin, Lawrence Livermore National Laboratories, and IN2P3, France. Operations are conducted by COO, IPAC, and UW.
This research is based on observations made with the Galaxy Evolution Explorer, obtained from the MAST data archive at the Space Telescope Science Institute, which is operated by the Association of Universities for Research in Astronomy, Inc., under NASA contract NAS 5--26555.
This publication makes use of data products from the Two Micron All Sky Survey, which is a joint project of the University of Massachusetts and the Infrared Processing and Analysis Center/California Institute of Technology, funded by the National Aeronautics and Space Administration and the National Science Foundation.
This publication makes use of data products from the Wide-field Infrared Survey Explorer, which is a joint project of the University of California, Los Angeles, and the Jet Propulsion Laboratory/California Institute of Technology, funded by the National Aeronautics and Space Administration.

YSE-PZ was developed by the UC Santa Cruz Transients Team. The UCSC team is supported in part by NASA grants NNG17PX03C, 80NSSC19K1386, and 80NSSC20K0953; NSF grants AST--1518052, AST--1815935, and AST--1911206; the Gordon \& Betty Moore Foundation; the Heising-Simons Foundation; a fellowship from the David and Lucile Packard Foundation to R.J.\ Foley; Gordon and Betty Moore Foundation postdoctoral fellowships and a NASA Einstein Fellowship, as administered through the NASA Hubble Fellowship program and grant HST-HF2-51462.001, to D.O.\ Jones; and an NSF Graduate Research Fellowship, administered through grant DGE--1339067, to D.A.\ Coulter.

\end{acknowledgments}

%

\vspace{5mm}
\facilities{{\it HST}(STIS), {\it Swift}(XRT and UVOT), AAVSO, CTIO:1.3m,
CTIO:1.5m,CXO}


\software{YSE-PZ was developed by the UC Santa Cruz Transients Team with support from The UCSC team is supported in part by NASA grants NNG17PX03C, 80NSSC19K1386, and 80NSSC20K0953; NSF grants AST-1518052, AST-1815935, and AST-1911206; the Gordon \& Betty Moore Foundation; the Heising-Simons Foundation; a fellowship from the David and Lucile Packard Foundation to R. J. Foley; Gordon and Betty Moore Foundation postdoctoral fellowships and a NASA Einstein fellowship, as administered through the NASA Hubble Fellowship program and grant HST-HF2-51462.001, to D. O. Jones; and a National Science Foundation Graduate Research Fellowship, administered through grant No. DGE-1339067, to D. A. Coulter.}



\appendix

\begin{table}\label{tab:phot}
    \centering
\caption{Photometry of SN\,2020acct. All reported magnitudes are in the AB system and corrected for foreground extinction.}
    \begin{tabular}{| cllll | cllll | cllll |}
    \hline
    Band & MJD & Phase & Mag. & Err. & Band & MJD & Phase & Mag. & Err. & Band & MJD & Phase & Mag. & Err. \\
    \hline
    UVW2 & 59197.85 & -54.66 & 19.01 & 0.10 &     PS i & 59258.46 & 3.92 & 19.00 & 0.04 &       P60 r & 59264.30 & 9.57  & 19.13 & 0.05 \\
    UVW2 & 59199.91 & -52.67 & 19.71 & 0.16 &     PS i & 59267.42 & 12.59 & 19.58 & 0.14 &      P60 r & 59266.26 & 11.46 & 19.18 & 0.08 \\
    UVW2 & 59203.88 & -48.82 & 20.28 & 0.20 &     PS z & 59192.65 & -59.68 & 19.57 & 0.10 &     P60 r & 59268.30 & 13.43 & 19.55 & 0.09 \\
    UVW2 & 59257.45 & 2.95  & 20.84 & 0.16 &      PS z & 59193.65 & -58.72 & 18.67 & 0.05 &     P60 r & 59272.28 & 17.28 & 20.04 & 0.30 \\
    UVW2 & 59260.52 & 5.91  & 20.55 & 0.20 &      PS z & 59195.65 & -56.78 & 18.20 & 0.03 &     P60 r & 59280.27 & 25.00 & 20.50 & 0.13 \\
    UVW2 & 59262.24 & 7.58  & 21.04 & 0.29 &      PS z & 59207.47 & -45.36 & 19.13 & 0.07 &     P60 r & 59292.23 & 36.57 & 21.12 & 0.24 \\
    UVW2 & 59268.62 & 13.74 & 20.39 & 0.24 &      PS z & 59212.59 & -40.41 & 19.94 & 0.11 &     P60 r & 59294.21 & 38.48 & 20.80 & 0.23 \\
    UVM2 & 59197.86 & -54.65 & 18.66 & 0.10 &     PS z & 59219.51 & -33.72 & 20.46 & 0.17 &     P60 r & 59306.24 & 50.10 & 21.38 & 0.28 \\
    UVM2 & 59199.91 & -52.67 & 19.34 & 0.13 &     PS z & 59220.67 & -32.60 & 20.33 & 0.15 &     c & 59205.53 & -47.23 & 18.88 & 0.05 \\
    UVM2 & 59201.24 & -51.38 & 19.60 & 0.19 &     PS z & 59237.46 & -16.37 & 20.67 & 0.29 &     c & 59207.53 & -45.30 & 19.17 & 0.07 \\
    UVM2 & 59203.89 & -48.82 & 20.07 & 0.17 &     PS y & 59194.67 & -57.73 & 18.60 & 0.10 &     c & 59229.54 & -24.03 & 20.54 & 0.20 \\
    UVM2 & 59257.46 & 2.96 & 20.53 & 0.15 &       PS y & 59206.68 & -46.12 & 19.30 & 0.13 &     c & 59251.51 & -2.80 & 18.72 & 0.05 \\
    UVM2 & 59260.52 & 5.92 & 20.46 & 0.24 &       PS w & 59206.50 & -46.30 & 18.81 & 0.01 &     c & 59253.50 & -0.87 & 18.60 & 0.04 \\
    UVM2 & 59262.24 & 7.58 & 20.45 & 0.20 &       PS w & 59255.39 & 0.96 & 18.86 & 0.04 &       c & 59255.41 & 0.98  & 18.64 & 0.05 \\
    UVM2 & 59268.62 & 13.74 & 20.50 & 0.24 &      P60 g & 59193.42 & -58.93 & 18.04 & 0.02 &    c & 59257.48 & 2.98  & 18.62 & 0.04 \\
    UVW1 & 59197.85 & -54.66 & 18.53 & 0.10 &     P60 g & 59198.39 & -54.13 & 17.96 & 0.02 &    c & 59261.42 & 6.78  & 18.89 & 0.05 \\
    UVW1 & 59199.90 & -52.67 & 18.93 & 0.14 &     P60 g & 59200.42 & -52.17 & 18.24 & 0.04 &    o & 59193.58 & -58.79 & 18.26 & 0.04 \\
    UVW1 & 59203.88 & -48.83 & 19.77 & 0.21 &     P60 g & 59203.41 & -49.28 & 18.79 & 0.04 &    o & 59208.52 & -44.34 & 18.84 & 0.08 \\
    UVW1 & 59262.24 & 7.58 & 20.21 & 0.25 &       P60 g & 59205.43 & -47.33 & 18.96 & 0.04 &    o & 59209.53 & -43.37 & 19.25 & 0.10 \\ 
    UVW1 & 59268.61 & 13.74 & 20.36 & 0.33 &      P60 g & 59225.46 & -27.97 & 21.30 & 0.32 &    o & 59217.59 & -35.58 & 19.61 & 0.20 \\
    U & 59197.85 & -54.66 & 17.95 & 0.09 &        P60 g & 59248.40 & -5.80 & 19.94 & 0.12 &     o & 59221.52 & -31.78 & 20.15 & 0.20 \\
    U & 59199.90 & -52.67 & 17.90 & 0.13 &        P60 g & 59250.30 & -3.96 & 19.00 & 0.05 &     o & 59249.64 & -4.60 & 19.09 & 0.18 \\
    U & 59203.88 & -48.83 & 18.65 & 0.17 &        P60 g & 59252.39 & -1.94 & 18.66 & 0.03 &     o & 59267.39 & 12.55 & 19.53 & 0.15 \\
    U & 59262.24 & 7.58 & 19.46 & 0.26 &          P60 g & 59254.29 & -0.11 & 18.58 & 0.03 &     &         &      &      &     \\
    B & 59197.85 & -54.66 & 17.65 & 0.10 &        P60 g & 59256.31 & 1.84 & 18.64 & 0.03 &      &         &      &      &     \\ 
    B & 59199.91 & -52.67 & 17.57 & 0.14 &        P60 g & 59258.36 & 3.83 & 18.75 & 0.03 &      &         &      &      &     \\ 
    B & 59201.24 & -51.38 & 17.44 & 0.17 &        P60 g & 59262.31 & 7.64 & 19.31 & 0.05 &      &         &      &      &     \\ 
    B & 59203.88 & -48.82 & 18.15 & 0.18 &        P60 g & 59264.33 & 9.60 & 19.40 & 0.05 &      &         &      &      &     \\ 
    B & 59262.24 & 7.58 & 18.37 & 0.20 &          P60 g & 59266.31 & 11.51 & 19.81 & 0.13 &      &         &      &      &     \\ 
    B & 59268.62 & 13.74& 18.13 & 0.21 &          P60 g & 59268.29 & 13.42 & 20.06 & 0.18 &      &         &      &      &     \\ 
    V & 59197.86 & -54.65& 17.40 & 0.16 &         P60 g & 59276.23 & 21.10 & 21.01 & 0.18 &      &         &      &      &     \\ 
    V & 59203.89 & -48.82& 17.30 & 0.20 &         P60 g & 59280.24 & 24.97 & 21.22 & 0.20 &      &         &      &      &     \\ 
    V & 59262.24 & 7.58 & 17.97 & 0.28 &          P60 r & 59193.49 & -58.87 & 18.31 & 0.03 &      &         &      &      &     \\ 
    V & 59268.62 & 13.74 & 17.87 & 0.32 &         P60 r & 59195.50 & -56.93 & 17.88 & 0.02 &      &         &      &      &     \\ 
    PS g & 59192.65 & -59.68 & 18.57 & 0.03 &     P60 r & 59198.47 & -54.05 & 18.04 & 0.02 &       &         &      &      &     \\ 
    PS g & 59200.58 & -52.02 & 18.30 & 0.02 &     P60 r & 59200.51 & -52.09 & 18.25 & 0.05 &       &         &      &      &     \\ 
    PS g & 59219.51 & -33.72 & 21.57 & 0.35 &     P60 r & 59203.47 & -49.22 & 18.62 & 0.04 &       &         &      &      &     \\ 
    PS g & 59256.34 & 1.88 & 18.65 & 0.04 &       P60 r & 59217.36 & -35.80 & 20.24 & 0.22 &      &         &      &      &     \\ 
    PS r & 59200.58 & -52.02 & 18.30 & 0.02 &     P60 r & 59219.36 & -33.86 & 20.62 & 0.25 &      &         &      &      &     \\ 
    PS r & 59207.47 & -45.36 & 19.09 & 0.04 &     P60 r & 59221.40 & -31.89 & 20.59 & 0.18 &      &         &      &      &     \\ 
    PS r & 59209.50 & -43.40 & 19.35 & 0.07 &     P60 r & 59229.41 & -24.15 & 20.70 & 0.29 &      &         &      &      &     \\ 
    PS r & 59246.37 & -7.76 & 20.07 & 0.26 &      P60 r & 59231.37 & -22.26 & 21.08 & 0.28 &      &         &      &      &     \\ 
    PS r & 59256.34 & 1.88 & 18.69 & 0.03 &       P60 r & 59250.34 & -3.92 & 18.95 & 0.06 &      &         &      &      &     \\ 
    PS i & 59207.51 & -45.32 & 18.99 & 0.04 &     P60 r & 59252.30 & -2.03 & 18.74 & 0.03 &      &         &      &      &     \\ 
    PS i & 59209.50 & -43.40 & 19.39 & 0.07 &     P60 r & 59254.39 & -0.01 & 18.67 & 0.03 &      &         &      &      &     \\ 
    PS i & 59221.49 & -31.81 & 20.91 & 0.17 &     P60 r & 59256.41 & 1.94 & 18.68 & 0.03 &      &         &      &      &     \\ 
    PS i & 59223.43 & -29.93 & 20.95 & 0.36 &     P60 r & 59258.33 & 3.79 & 18.80 & 0.05 &     &         &      &      &     \\ 
    PS i & 59246.37 & -7.76  & 20.75 & 0.23 &     P60 r & 59262.24 & 7.57 & 19.09 & 0.07 &      &         &      &      &     \\ 
    \hline
    \end{tabular}
\end{table}

\begin{table}\label{tab:spec}
    \centering
\caption{Spectroscopic observations of SN\,2020acct. All reported phases are with respect to maximum light of the second peak.}
    \begin{tabular}{|clllcl|}
    \hline
    MJD & Phase (days) & Telescope & Instrument & Range ({\AA}) & Exp. Time  \\
    \hline
    59194 &  -58 & Keck-I & LRIS & 3600-9670 &  750s\\
    59197 &  -54 & NOT & ALFOSC & 3620-8575  &  900s\\
    59254 &  0 & NOT & ALFOSC & 3620-8575  &  900s\\
    59257 &  +3 & Keck-I & LRIS & 3600-9670 & 900s \\
    59264 &  +9 & Shane & Kast & 3130-9615  & 3600s \\
    59312 &  +56 & Keck-I & LRIS & 3600-9670 &  1200s\\
    59346 &  +89 & Keck-I & LRIS & 3600-9670 & 1800s \\
     
    \hline
    \end{tabular}
\end{table}

\begin{table}\label{tab:mosfit}
    \centering
\caption{{\tt{MOSFiT}} parameters for SN\,2020acct. The two peaks are fit independently, the first peak with the CSM model and the second with the Arnett model.}
    \begin{tabular}{|lllll|}
    \hline
    {\bf{Peak 1}} &(CSI, S=0)&&& \\
    Param. & Units & Prior&  Range & Output \\ 
    \hline
    $t_{0}$ & days  & Flat & -8.5 - 0 &-2.76$^{+0.50}_{-0.60}$ \\ 
    $M_{CSM}$ & M$_{\odot}$ & Log-flat & 0.01 - 10  & 0.046$^{+0.014}_{-0.014}$\\
    $R_{CSM}$ & cm & Log-flat & 10$^{8}$ - 10$^{16}$ & 3.05$^{+7.8}_{-2.5}$\,$\times10^{12}$\\ 
    $M_{ej}$ &  M$_{\odot}$& Log-flat & 0.001 - 10 & 0.6$^{+3.6}_{-0.4}$\,$\times10^{-2}$ \\ 
    $v_{ej}$ & km\,s$^{-1}$ & Log-flat & 1 - 100 & 15.1$^{+1.8}_{-1.6}$\,$\times10^{3}$  \\ 
    $\rho$ &  g\,cm$^{-3}$& Log-flat &  10$^{-10}$ - 10$^{8}$ & 0.13$^{+0.11}_{-0.06}$\,$\times10^{-8}$\\ 
    T$_{min}$ & K & Log-flat & 80 - 25000 & $1.07^{+0.13}_{-0.13}$\,$\times10^{4}$\\ 
    \hline
    {\bf{Peak 1}} &   (CSI, S=2)&&& \\
    Param. & Units & Prior & Range & Output \\ 
    \hline
    $t_{0}$ & days  & Flat & -8.5 - 0 &-3.46$^{+0.33}_{-0.59}$ \\ 
    $M_{CSM}$ & M$_{\odot}$ & Log-flat & 0.01 - 10 & 0.05$^{+0.03}_{-0.01}$\\
    $R_{CSM}$ & cm & Log-flat & 10$^{8}$ - 10$^{16}$ & 16.0$^{+15}_{-7}$\,$\times10^{12}$ \\   
    $M_{ej}$ &  M$_{\odot}$& Log-flat & 0.001 - 10 & 0.07$^{+2.9}_{-0.50}$\,$\times10^{-2}$ \\ 
    $v_{ej}$ & km\,s$^{-1}$ & Log-flat & 1 - 100 & 12.5$^{+1.5}_{-1.30}$\,$\times10^{3}$ \\ 
    $\rho$ & g\,cm$^{-3}$& Log-flat &  10$^{-10}$ - 10$^{8}$ & 0.5$^{+0.31}_{-0.27}$\,$\times10^{-8}$\\ 
    T$_{min}$ &  K & Log-flat & 80 - 25000 & 1.07$^{+0.13}_{-0.11}$\,$\times10^{4}$\\ 
    \hline
    {\bf{Peak 1}} &   (CSI, S free)&&& \\
    Param. & Units & Prior & Range & Output \\ 
    \hline
    $t_{0}$ & days  & Flat & -8.5 - 0 &-2.16$^{+0.39}_{-0.51}$\\ 
    $M_{CSM}$ & M$_{\odot}$ & Log-flat & 0.01 - 10 &0.01$^{+0.004}_{-0.005}$\\ 
    $R_{CSM}$ & cm & Log-flat & 10$^{8}$ - 10$^{16}$ &0.31$^{+0.34}_{-0.12}$\,$\times10^{12}$\\ 
    $M_{ej}$ &  M$_{\odot}$ & Log-flat & 0.001 - 10 &0.009$^{+0.025}_{-0.005}$\,$\times10^{-2}$\\ 
    $v_{ej}$ & km\,s$^{-1}$ & Log-flat & 1 - 100 & 18.2$^{+2.2}_{-1.9}$\,$\times10^{3}$\\ 
    $\rho$ &  g\,cm$^{-3}$& Log-flat &  10$^{-10}$ - 10$^{8}$ &9.12$^{+24}_{-17}$\,$\times10^{-8}$\\ 
    T$_{min}$ &  K & Log-flat & 80 - 25000 &1.07$^{+0.07}_{-0.07}$\,$\times10^{4}$\\ 
    s$_{min}$ &   & Flat & 0 - 2 &$0.84^{+0.30}_{-0.27}$\\ 
    \hline
    {\bf{Peak 2}}&  (Ni$^{56}$)&&& \\
    Param. & Units & Prior & Range & Output \\ 
    \hline
    $t_{0}$ & days & Flat & -80 - 0 & -3.08$^{+0.28}_{-0.33}$ \\
    $f_{Ni}$ & - & Log-flat & 10$^{-3}$ - 1.0  & 0.91 $^{+0.07}_{-0.11}$ \\ 
    $M_{ej}$ & M$_{\odot}$ & Log-flat & 0.01 - 100  & 0.132$^{+0.011}_{-0.008}$ \\ 
    $v_{ej}$ & km\,s$^{-1}$ & Gaussian & $\overline{v_{ej}}$=5.9  & 6.23$^{+0.22}_{-0.20}$\,$\times10^{3}$ \\ 
     &  &  & $\sigma$=0.7  &  \\ 
    $T_{min}$ & K & Log-flat & 80 - 25000 & 5.25$^{+1.4}_{-0.4}$\,$\times10^{4}$  \\ 
    \hline
    \end{tabular}
\end{table}

\begin{figure*}
    \centering
\includegraphics[width=0.73\textwidth]{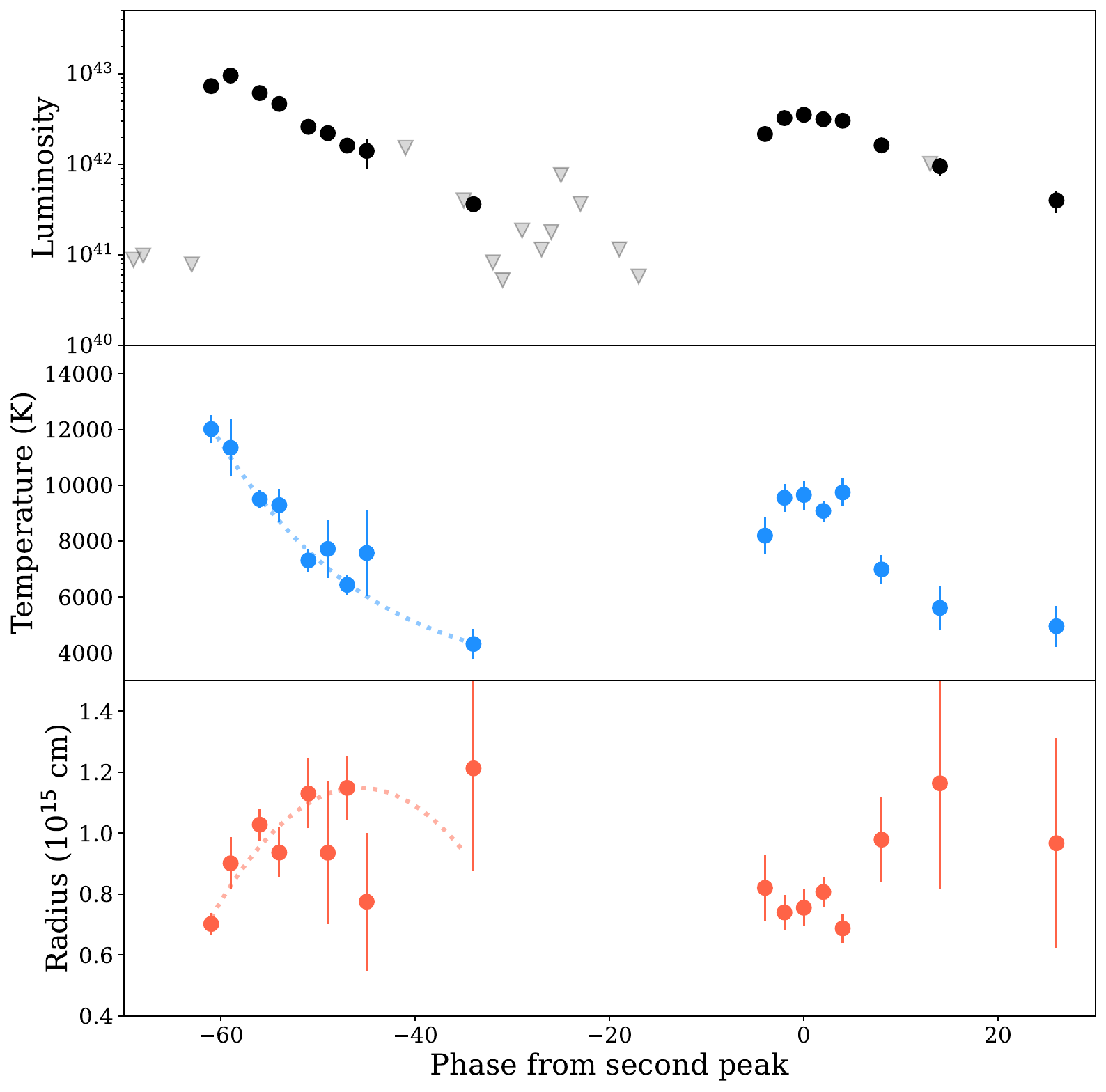}
\caption{Luminosity, temperature and radial evolution of the blackbody fits to the two peaks of SN\,2020acct. Phases are given with respect to the second peak.}
\label{extfig:BB_fit}
\end{figure*}

\begin{figure*}
\centering
\includegraphics[width=0.7\textwidth]{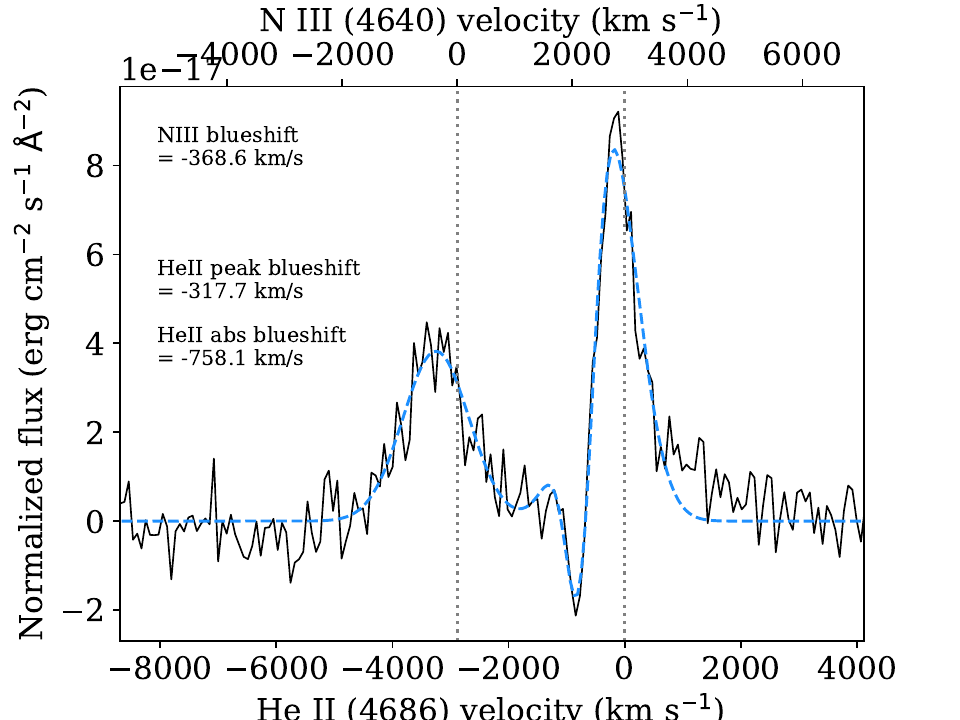}
\caption{Modelling the emission profiles around $\lambda$4650 as \ion{He}{2} and \ion{N}{3} . To recreate the profile, an additional \ion{He}{2} absorption component was included. The profiles of both emission lines are narrow (900\,km\,s$^{-1}$ for \ion{N}{3} and 450\,km\,s$^{-1}$ for \ion{He}{2}), and are offset by $>$300\,km\,s$^{-1}$ from the rest frame. As narrow lines typically indicate the presence of low density/optically thin material, the only way to produce this blueshift would be through attenuation of the red wing of the line through dust condensation. }
 \label{fig:heiiniii}
\end{figure*}

\begin{figure*}
\includegraphics[width=\textwidth]{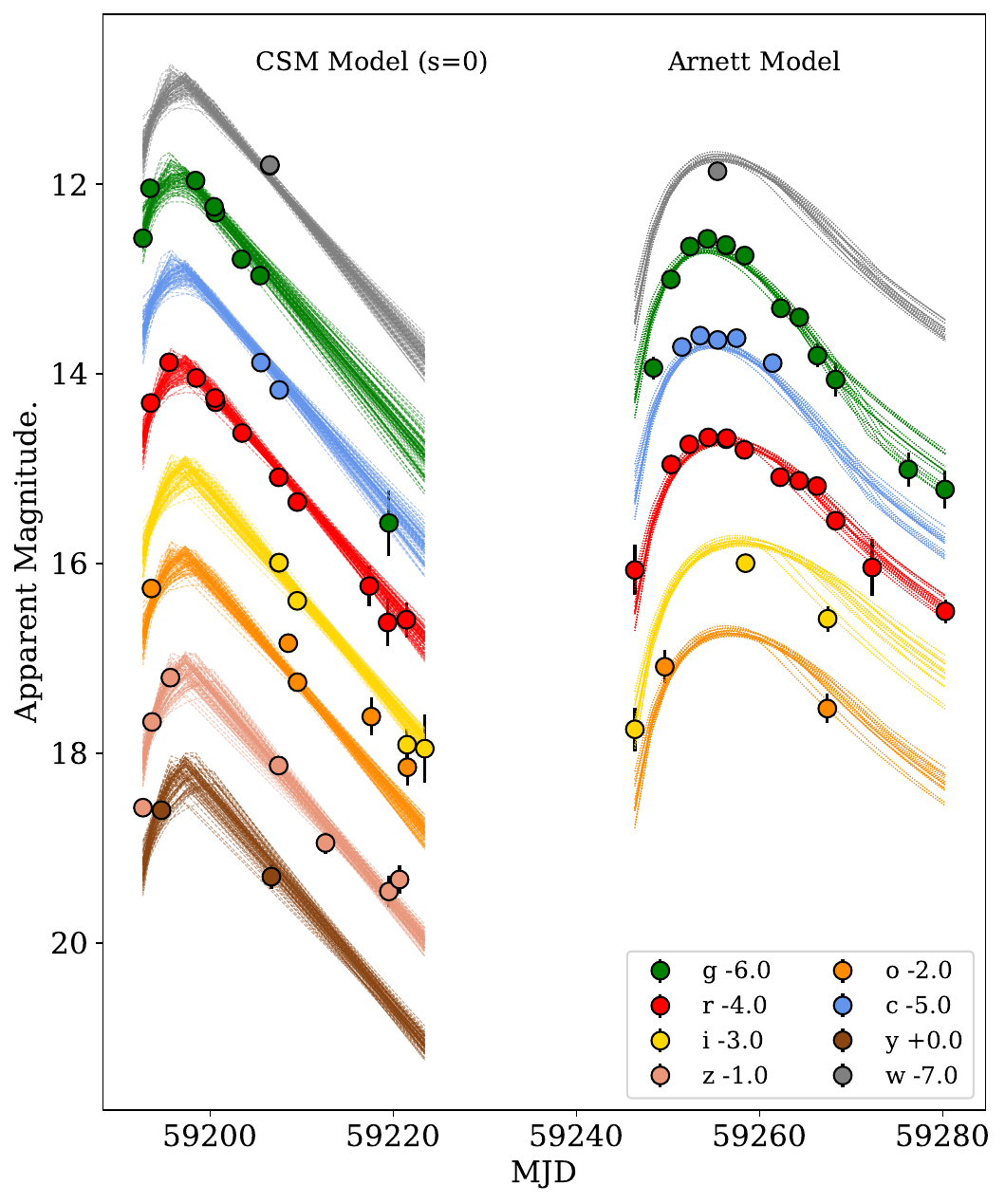}
\caption{{\tt{MOSFiT}} modelling of SN\,2020acct, fitting the first peak with CSM interaction from \citealt{Chatzopolos2012} (here showing the output for shell-like CSM, $S=0$), and the second peak with \citealt{Arnett1982} Ni$^{56}$ decay. Faded triangles are photometric upper limits.}
 \label{fig:mosfit}
\end{figure*}

\begin{figure*}
\includegraphics[width=\textwidth]{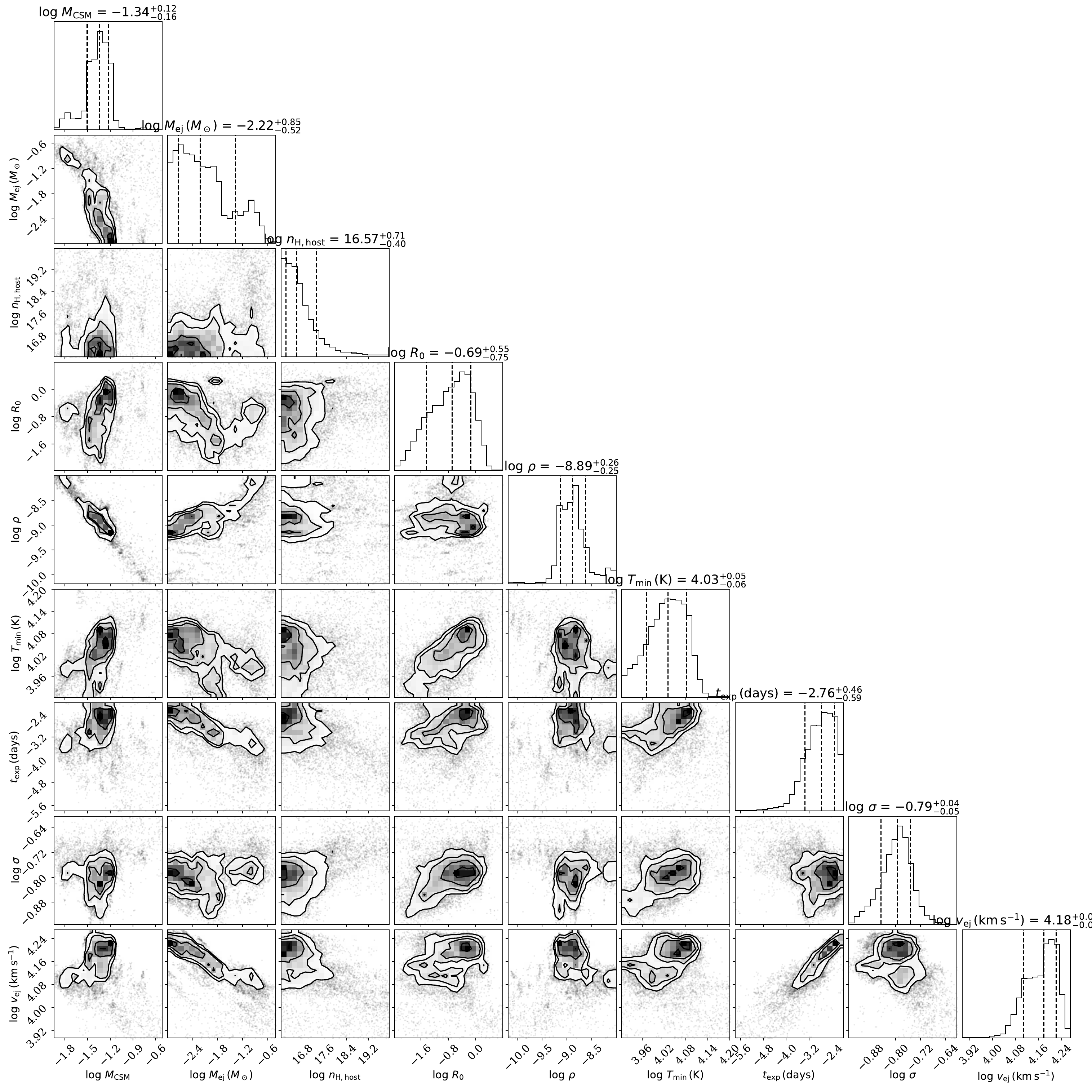}
\caption{{\tt{MOSFiT}} parameter distributions to CSM interaction modelling of 1st peak of SN\,2020acct with fixed $s=0$. We note the minimal change to fit quality when setting $s=2$ or leaving it as a free parameter. Our resulting parameters are given in Table \ref{tab:mosfit}.}
\label{extfig:mosfit_peak1_s0}
\end{figure*}



\begin{figure*}
\includegraphics[width=\textwidth]{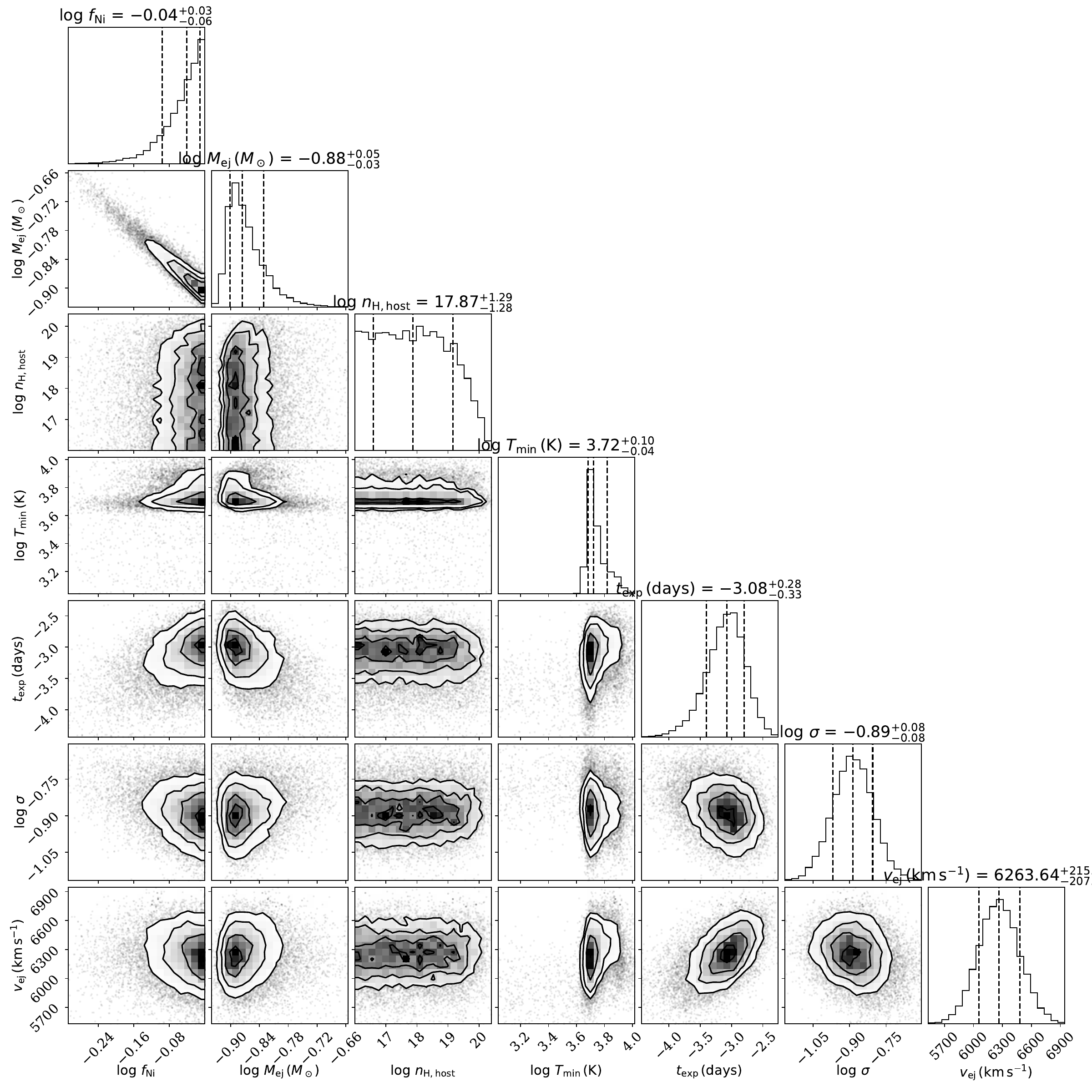}
\caption{{\tt{MOSFiT}} parameter distributions to Ni$^{56}$ powered light curve model to 2nd peak of SN\,2020acct}
\label{extfig:mosfit_peak2}
\end{figure*}

\begin{figure*}
\includegraphics[width=0.49\textwidth,trim=2.cm 1.5cm 2.cm 2.5cm,clip]{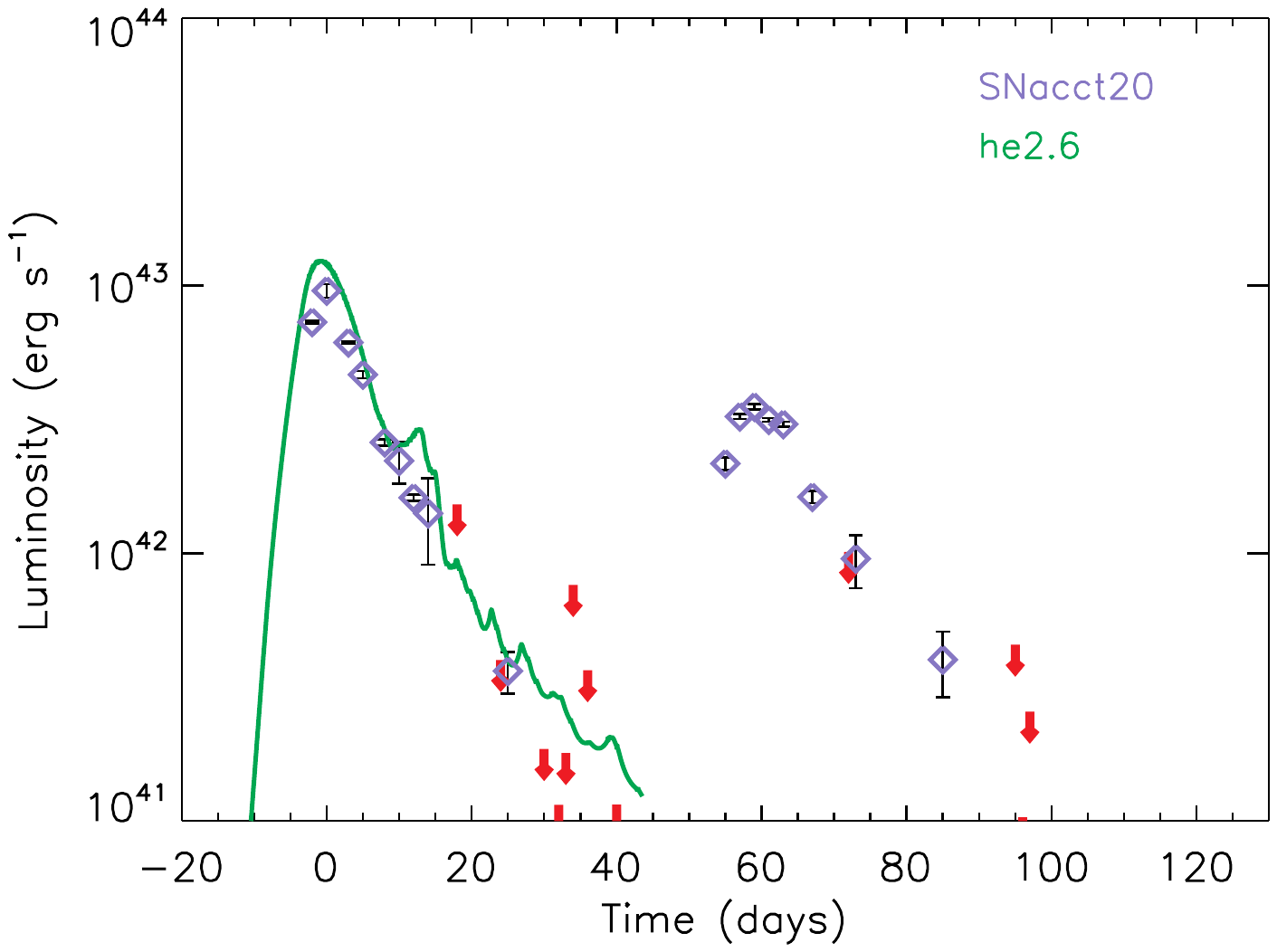}
\includegraphics[width=0.49\textwidth,trim=2.cm 1.5cm 2.cm 2.5cm,clip]{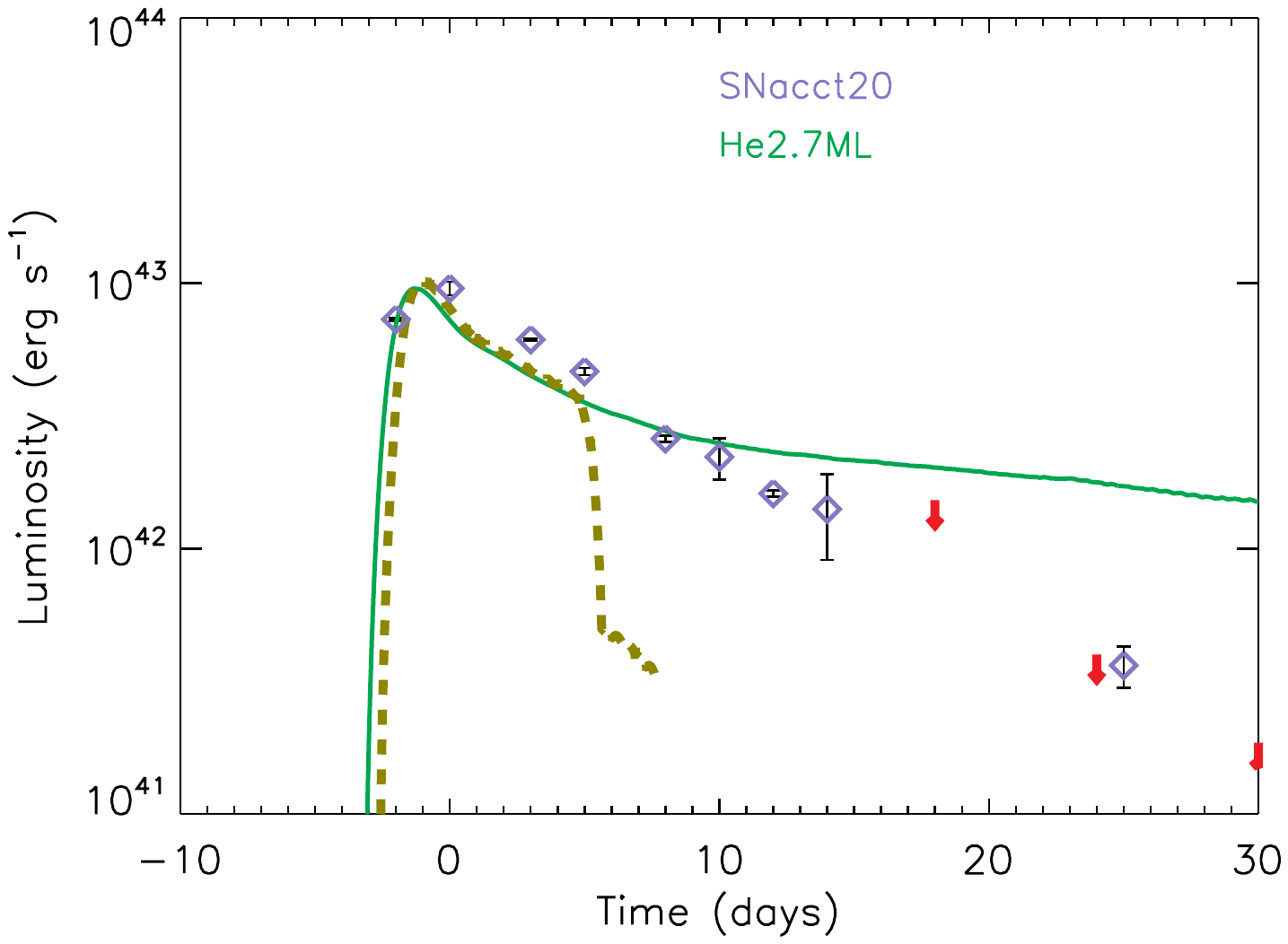}
\caption{Two low-mass helium star models for the first peak of SN 2020acct. {\textit{Left:}} A model that experienced a silicon flash 101 days before exploding with a very low energy, $2.3 \times 10^{49}$ erg. The light curve is produced by the explosion impacting the shell ejected by the flash. The initial helium star mass was 2.6\,M$_{\odot}$ and the pre-SN mass, 2.14\,M$_{\odot}$, of which the flash ejected 0.47\,M$_{\odot}$ with energy $6\times 10^{47}$\,erg. The first peak is fit well by this model, but the required explosion is low for neutrino-powered SNe in this mass range, and there is no obvious mechanism for making the second peak. {\textit{Right:}} A similar model without a silicon flash that interacted with a steady wind of 0.1\,M$_{\odot}$\,yr$^{-1}$ with velocity 1000\,km\,s$^{-1}$. The initial helium core mass here was 2.7\,M$_{\odot}$; the pre-SN mass 2.21\,M$_{\odot}$; and the explosion kinetic energy, $9\times 10^{49}$\,erg. The initial peak is fit well, but the model declines too slowly after the peak.  This could indicate decreasing optical efficiency for the shock, which lies above the photosphere (dashed line), or a non-steady wind with density declining faster than r$^{-2}$.}
\label{extfig:helowmass}
\end{figure*}

\begin{figure*}
\centering
\includegraphics[width=0.49\textwidth,trim=2.cm 1.5cm 2.cm 2.5cm,clip]{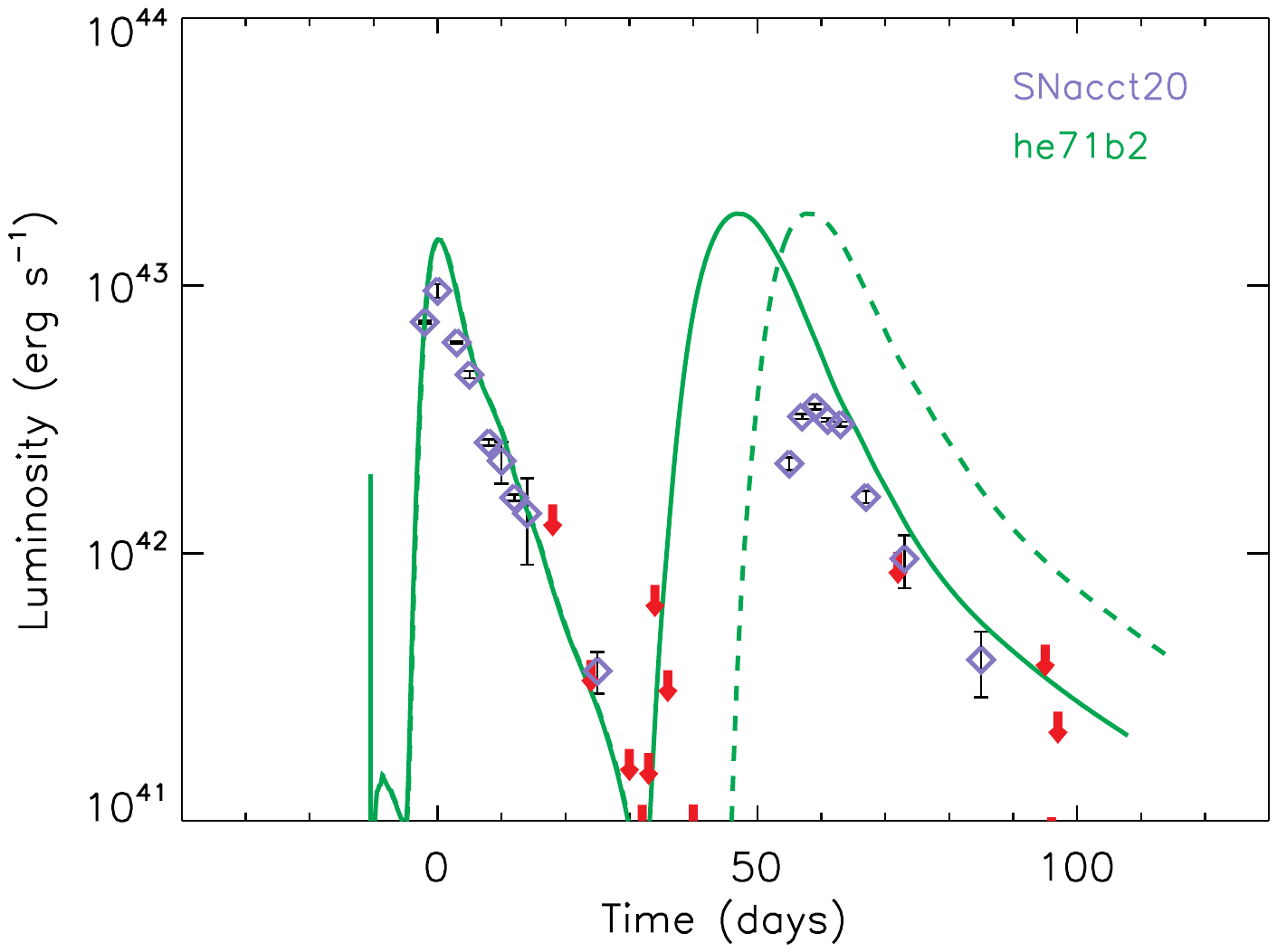}
\includegraphics[width=0.49\textwidth,trim=2.cm 1.5cm 2.cm 2.5cm,clip]{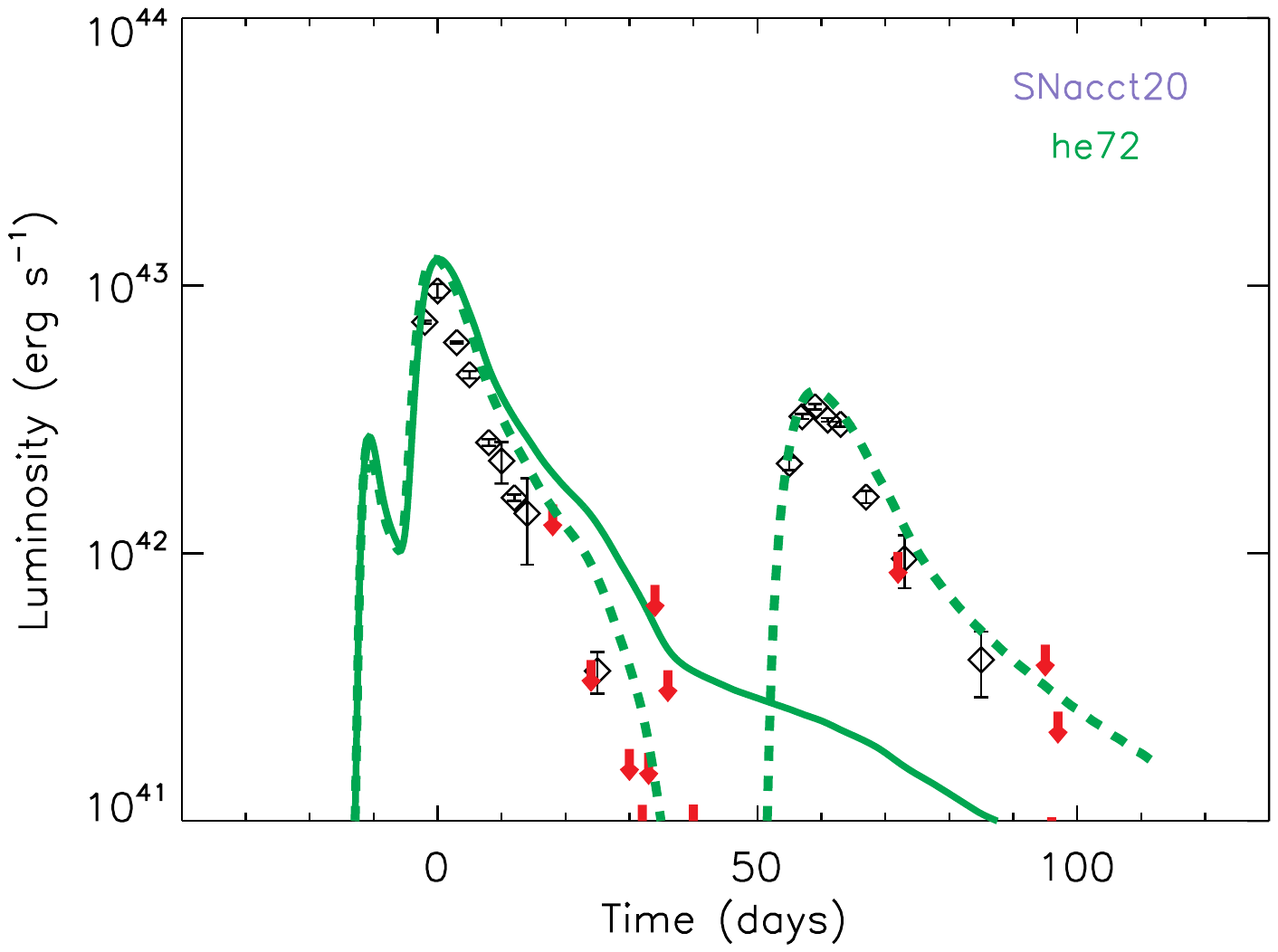}
\caption{Bolometric light curve of SN\,2020acct (points) compared to two standard PPISN models and modified models. {\textit{Left:}} Impact of increasing neutrino losses. The dashed line here shows the standard model of a 71 M$_{\odot}$ helium star (mass at the onset of PPI of 52 M$_{\odot}$) which undergoes 4 pulses prior to collapse. The solid line shows the modified model in which neutrino losses have been increased by 30\%. Increasing neutrino loses reduces the central temperature of the core, shortening the time delay between peaks. {\textit{Right:}} Impact of increased carbon abundance. With the solid line we show a 72 M$_{\odot}$ helium star (mass at the point of PPI of 52.92 M$_{\odot}$) with slightly enhanced carbon abundance but no neutrino losses. Despite having 4 pulses, the model generates only a single peak. Including both enhanced carbon and neutrino loses (dashed line), a weak fifth pulse makes the second peak of the light curve. See Section \ref{sec:PPImodel} for more details.}
\label{extfig:PPIhe71}
\end{figure*}

\begin{figure*}
\centering
\includegraphics{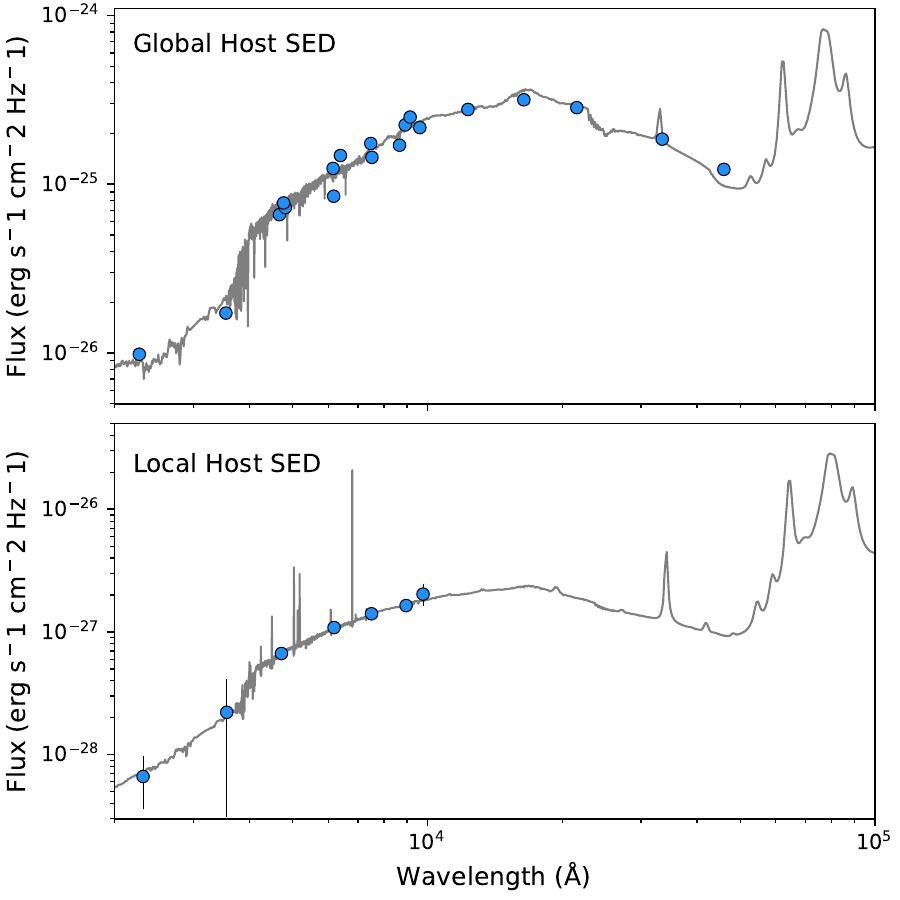}
\caption{{\tt{Prospector}} Spectral Energy Distribution fitting the global (top panel) and local (bottom panel) host photometry within 1`` of the explosion site of SN\,2020acct. Fitting the global SED, we find a stellar mass of $\log{(M_{\star} / M_{\odot})}=11.29^{+0.04}_{-0.06}$, and a current star formation rate of $(SFR / M_{\odot} yr^{-1}) = 17^{+15}_{-11}$. For the stellar population we find a
star formation rate of $\log{(SFR / M_{\odot} {\rm yr}^{-1})} = -1.1^{+0.9}_{-0.7}$ and average stellar age of $t_{\mathrm{age}}$ = 9.13$^{+4.5}_{-3.3}$ Myr.}
\label{extfig:SEDfits}
\end{figure*}

\bibliography{SN2020acct}{}
\bibliographystyle{aasjournal}



\end{document}